\documentclass[twocolumn,secnumarabic,amssymb, nolongbibliography, aps, prl]{revtex4-2}

\usepackage{graphicx,bm}
\usepackage{amsmath}
\usepackage{amssymb}
\usepackage{bm}
\usepackage{xcolor}
\usepackage{graphicx}  
\usepackage{dcolumn}   
\usepackage{bm}        
\usepackage{verbatim}   
\usepackage{graphicx}
\usepackage{epstopdf}
\usepackage{footnote}
\usepackage{epsfig}
\usepackage{subfigure}
\usepackage{amssymb}
\usepackage{amsmath,bm}
\usepackage{xcolor}
\usepackage{float}
\usepackage{subfigure}
\usepackage[most]{tcolorbox}
\usepackage{soul}
\usepackage{color}
\usepackage[colorlinks=true,linkcolor=blue,allcolors=blue]{hyperref}%
\usepackage[framemethod=TikZ]{mdframed}
\usepackage{lipsum}
\mdfdefinestyle{MyFrame}{%
	linecolor=blue,
	outerlinewidth=2pt,
	roundcorner=20pt,
	innertopmargin=\baselineskip,
	innerbottommargin=\baselineskip,
	innerrightmargin=20pt,
	innerleftmargin=20pt,
	backgroundcolor=gray!50!white}

\begin{document}
	
	{}
	
	\headsep = 40pt
	\title{Nonreciprocal Radiation in Time-Periodic Topological Quadrupole Lattices}
	\author{Sajjad Taravati and Jingbang Liu}
	
	\affiliation{School of Electronics and Computer Science, University of Southampton, Southampton SO17 1BJ, UK}
	\email{e-mail: s.taravati@soton.ac.uk}
	
\begin{abstract}
We realize a time-periodic quadrupole higher-order topological lattice and demonstrate its nonreciprocal radiation. A relative phase between two modulated coupling channels acts as a synthetic gauge field that breaks time-reversal symmetry without closing the bulk gap, and a numerical Floquet analysis finds that zero-quasienergy corner modes persist. With the corners coupled to free space, a signal injected at one corner is up-converted to the first Floquet sideband and radiated, and a free-space wave at that sideband is delivered preferentially to a second corner, with more than 21 dB contrast relative to the first, combining frequency conversion with nonreciprocal routing. The same radiated frequency is reached with modulation frequencies differing by more than an order of magnitude.
\end{abstract}
	
\maketitle

\emph{Introduction.} Higher-order topological insulators (HOTIs) extend the bulk-boundary correspondence to boundaries of codimension two or more, where corner or hinge states are protected by crystalline symmetry~\cite{Benalcazar2017a,Benalcazar2017b}. The quantized electric quadrupole insulator is the paradigmatic case: a synthetic $\pi$ flux through every plaquette makes the two mirror symmetries anticommute, quantizes the bulk quadrupole moment to $q_{xy}=e/2$, and pins fractional charges $\pm e/2$ to the corners~\cite{Benalcazar2017a,Benalcazar2017b}, as observed in microwave and photonic lattices~\cite{peterson2018quantized,schulz2022photonic}.

Periodic driving enriches this picture. A Floquet topological phase can host boundary modes at both quasienergy $0$ and $\pi/T$, and the $\pi$ modes have no static counterpart~\cite{Rudner2013,Oka2019}. Floquet HOTIs are predicted to host corner modes at both quasienergies~\cite{Peng2019,Huang2020,NagJuricicRoy2019}, and time-periodic corner states have been observed in an acoustic lattice, where a spatial axis emulates time~\cite{Zhu2022}, and in space-time circuit networks~\cite{Qian2026}. Quadrupole lattices realized in microwave and photonic platforms are passive and guided-wave or near-field coupled~\cite{peterson2018quantized,schulz2022photonic}, and the time-periodic demonstrations above do not couple the corners to free-space radiation. To our knowledge, a time-periodic quadrupole lattice whose corners radiate into free space has not been reported.

Independently, reciprocity is a fundamental constraint on linear, time-invariant electromagnetic systems: in the absence of a symmetry-breaking bias, forward and backward transmission between any two ports must be identical~\cite{taravati_PRApp_2019}. Breaking this constraint --- to build isolators~\cite{taravati2017self,Taravati_PRB_2017,Taravati_PRAp_2018,Taravati_AMTech_2021}
circulators~\cite{dinc2017millimeter,taravati2022low,wu2024analysis} --- has traditionally required magneto-optic materials biased by a static magnetic field, a route that is bulky, lossy, and difficult to integrate~\cite{Lax_1962,fay1965operation,fan2012magnetically}. Space-time and time modulation techniques offer magnet-free alternatives: a coupling that is modulated periodically in time imparts opposite phases to waves propagating in opposite directions. Such a technique produces a synthetic gauge field and hence an asymmetric effective coupling without magnetic bias~\cite{Taravati_Kishk_MicMag_2019,taravati2025designing,taravati20234d,ptitcyn2026photonic,patel2026photonic,Taravati_PRB_Mixer_2018,zang2019nonreciprocal,Taravati_PRAp_2018,taravati2020full,Taravati_AMTech_2021,fang2012photonic}, yielding amplification~\cite{Tien_JAP_1958,taravati2026temporal},
multifunctional metasurfaces and antennas~\cite{taravati2020full,Taravati_AMA_PRApp_2020,wu2025space,taravati2025_entangle,Taravati_ACSP_2022,taravati2025light}, unidirectional beam-splitting~\cite{Taravati_Kishk_PRB_2018}, and frequency conversion~\cite{Taravati_PRB_Mixer_2018,taravati2021pure,taravati2026BraggConv}.

Here we introduce radiating, time-periodic quadrupole lattices. In a quadrupole lattice driven so that two coupling channels are modulated with a relative phase, that phase acts as a synthetic gauge field that breaks time-reversal symmetry while the bulk gap stays open. We realize this at microwave frequencies, in a time-modulated metasurface whose corners connect to free space through dipole antennas, so that the corners act as transceivers. A signal at the transmit (TX) corner is up-converted to the first Floquet sideband and radiated, and a free-space wave at that sideband is delivered preferentially to the receive (RX) corner, with more than $21$~dB contrast relative to TX. Frequency conversion and nonreciprocal routing are then carried by the same resonance rather than by separate components.

\emph{Driven quadrupole lattice.} The static quadrupole insulator is a lattice of four-site plaquettes with intra-cell coupling $\gamma$, inter-cell coupling $\lambda$, and a pattern of negative bonds such that every plaquette, within and between cells, carries flux $\pi$. Its Bloch Hamiltonian is~\cite{Benalcazar2017a}
\begin{equation}
\begin{split}
H(\mathbf{k})=&[\gamma+\lambda\cos k_x]\Gamma_4+\lambda\sin k_x\,\Gamma_3\\
&+[\gamma+\lambda\cos k_y]\Gamma_2+\lambda\sin k_y\,\Gamma_1,
\end{split}
\label{eq:Hk}
\end{equation}
with $\Gamma_i$ the $4\times4$ gamma matrices of Ref.~\cite{Benalcazar2017a}. The corner-mode phase occurs for $|\gamma/\lambda|<1$. The real-space unit-cell matrix and the microwave realization of the $\pi$ flux are in the Supplemental Material (SM)~\cite{SM}.

\begin{figure}
	\begin{center}
\subfigure[]{\label{Fig:1a}
	\includegraphics[width=0.8\columnwidth]{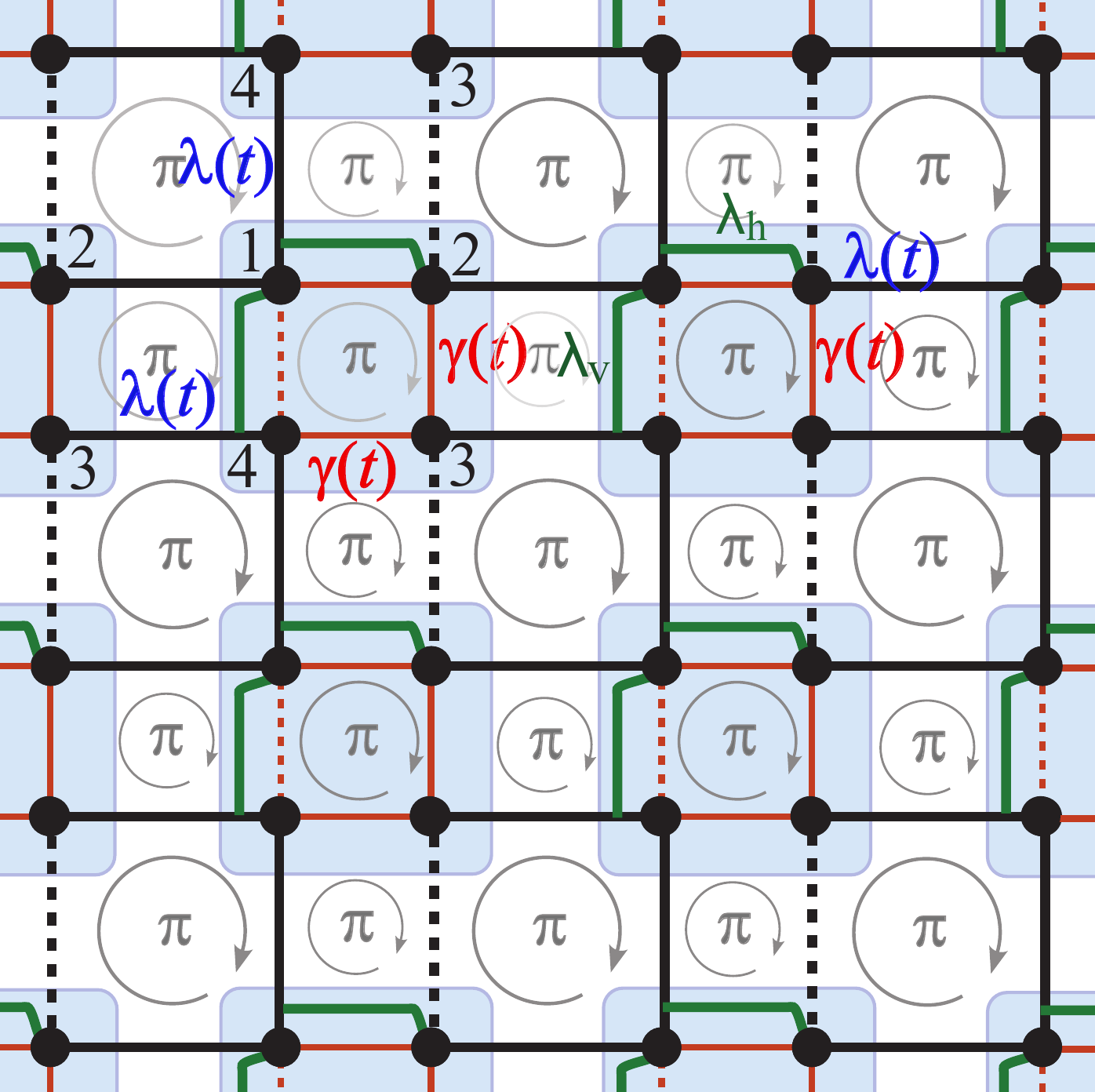}}
\subfigure[]{\label{Fig:1b}
	\includegraphics[width=1\columnwidth]{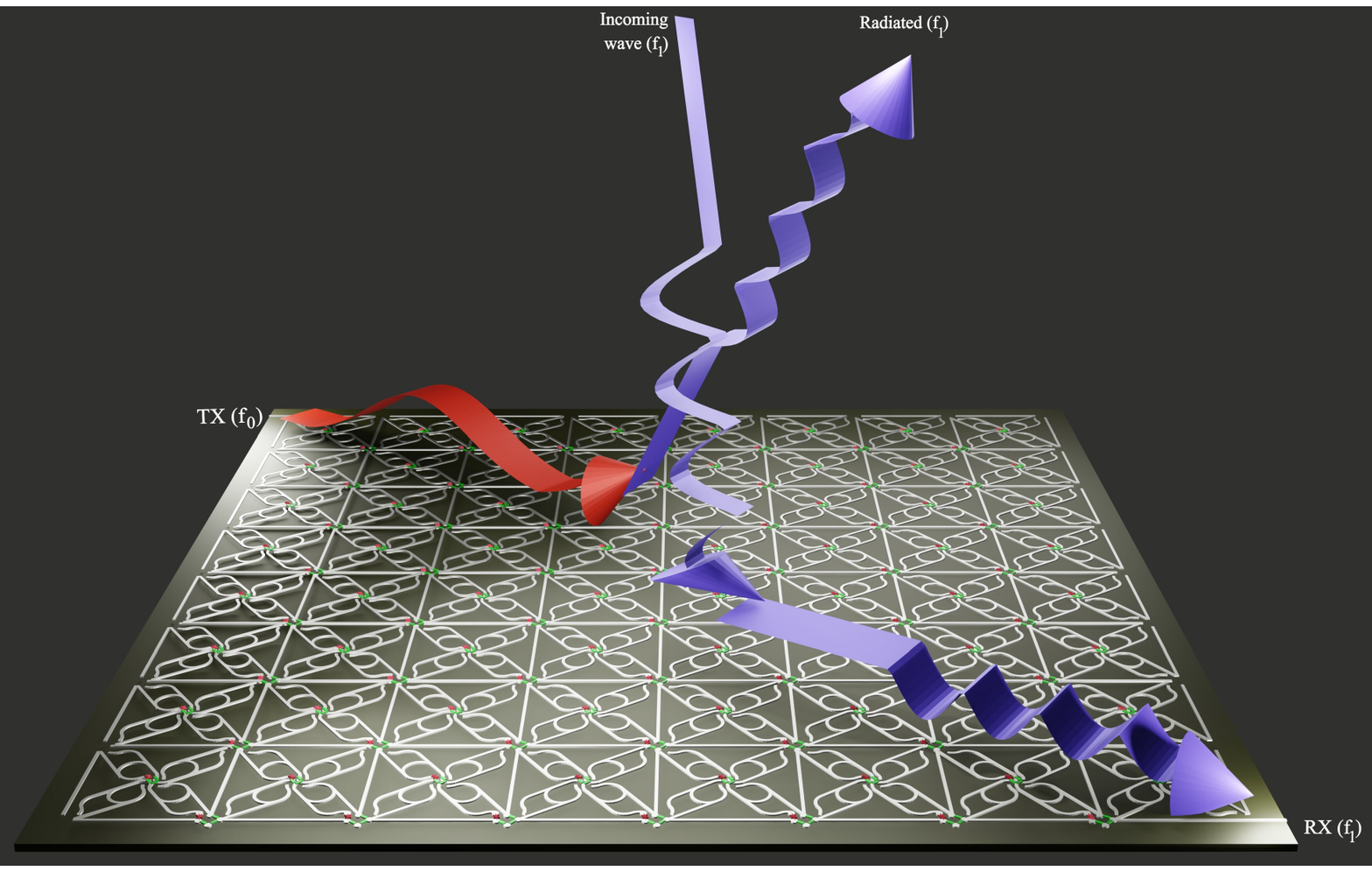}}
\caption{(a)~Driven quadrupole lattice. Tight-binding lattice with the time-dependent couplings $\gamma(t)$ and $\lambda(t)$ (\eqref{eq:drive}) and the static dipole couplings $\lambda_{\rm v}$ and $\lambda_{\rm h}$. Dotted bonds are negative, so that every plaquette, within and between unit cells, carries a synthetic $\pi$ flux. (b)~Circulator operation. A signal at $f_0$ injected at TX (top-left corner, $R_1$) is up-converted by the modulated array and radiated at $f_1=f_0+ f_\text{m}$. A free-space wave at $f_1$ is captured and delivered, at the same frequency and without further conversion, to RX (bottom-right corner, $R_3$), while the reverse itinerary is isolated.}
\label{fig:1}
	\end{center}
\end{figure}

We drive the lattice with two time-periodic couplings (Fig.~\ref{fig:1}a),
\begin{subequations}
\label{eq:drive}
\begin{align}
\gamma(t)&=\gamma_0+\delta\gamma\cos\Omega t,\\
\lambda(t)&=\lambda_0+\delta\lambda\sin\Omega t,
\end{align}
\end{subequations}
with $\Omega=2\pi f_\text{m}$ and $T=2\pi/\Omega$, and $H(\mathbf{k},t)$ follows from \eqref{eq:Hk} by $\gamma\to\gamma(t)$, $\lambda\to\lambda(t)$. The cosine-versus-sine offset means that no time shift maps the drive onto its time reverse, so the drive itself lacks time-reversal symmetry. Its microscopic realization, as modulation phases of the varactors, is described below.

Because $H(\mathbf{k},t+T)=H(\mathbf{k},t)$, the dynamics are governed by the Floquet operator $U(T)$, with $H_F=(i/T)\log U(T)$ and quasienergies $\varepsilon\in[-\pi/T,\pi/T]$~\cite{Goldman2015}. A Floquet phase can host corner modes at $\varepsilon=0$ and, in anomalous phases, at $\varepsilon=\pi/T$~\cite{Rudner2013,Peng2019,Huang2020}. We checked numerically (SM) that for the drive of \eqref{eq:drive} and the couplings of our array, the zero-quasienergy corner modes of the static lattice persist, with corner weight $0.76$ to $0.91$ on a finite lattice at the modulation frequencies used. The same calculation finds no $\pi$-quasienergy corner modes at these parameters, and we make no claim about them.

The instantaneous gap closes when $|\gamma(t)/\lambda(t)|=1$, which has real solutions only if $\lambda_0-\gamma_0<\sqrt{\delta\gamma^2+\delta\lambda^2}$ (SM). We operate in the opposite limit (Regime~I),
\begin{equation}
\lambda_0-\gamma_0>\sqrt{\delta\gamma^2+\delta\lambda^2},
\label{eq:regime}
\end{equation}
in which $|\gamma(t)/\lambda(t)|<1$ throughout the cycle and the instantaneous Hamiltonian never leaves the corner-mode phase. For our array, $\gamma_0=35$~MHz, $\lambda_0=150$~MHz, $\delta\gamma\approx10$~MHz and $\delta\lambda\approx20$~MHz, so $\lambda_0-\gamma_0=115$~MHz exceeds $\sqrt{\delta\gamma^2+\delta\lambda^2}\approx22.4$~MHz and the device operates in Regime~I.

\begin{figure}
	\subfigure[]{\label{Fig:3a}
		\includegraphics[width=0.55\columnwidth]{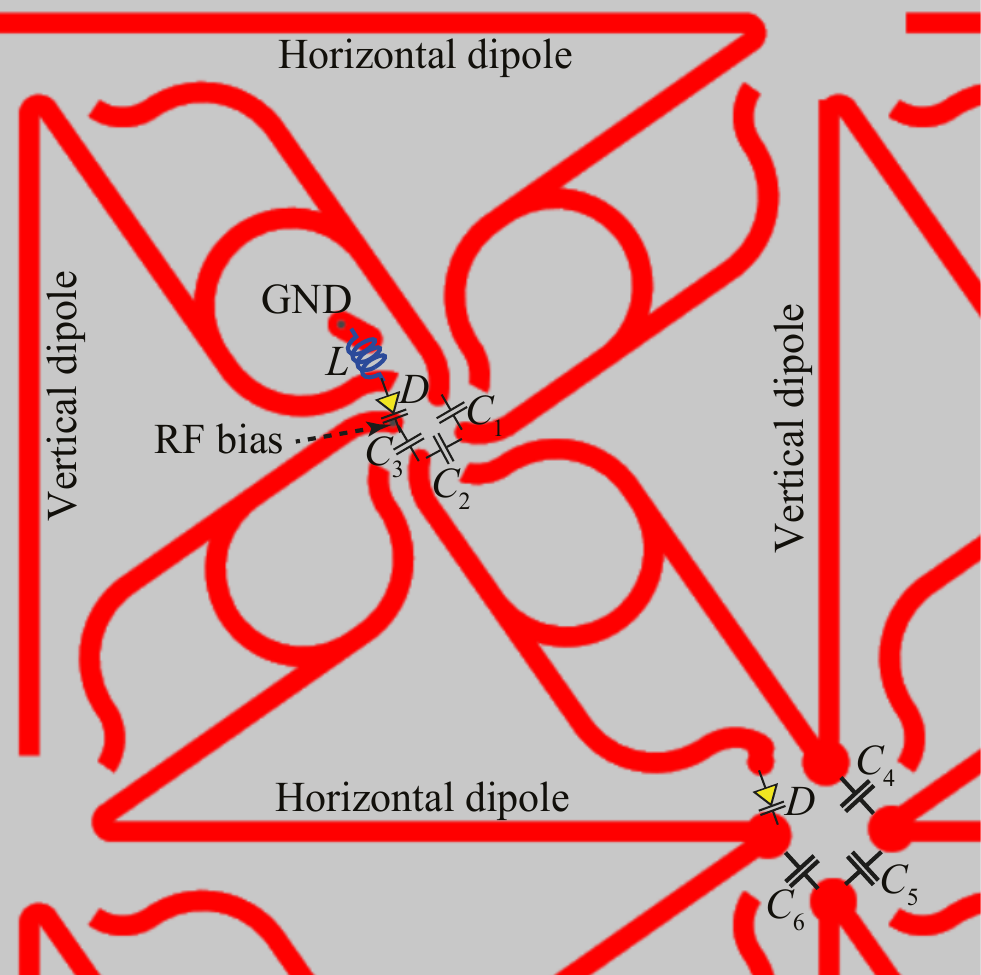}}
	\subfigure[]{\label{Fig:3b}
		\includegraphics[width=0.38\columnwidth]{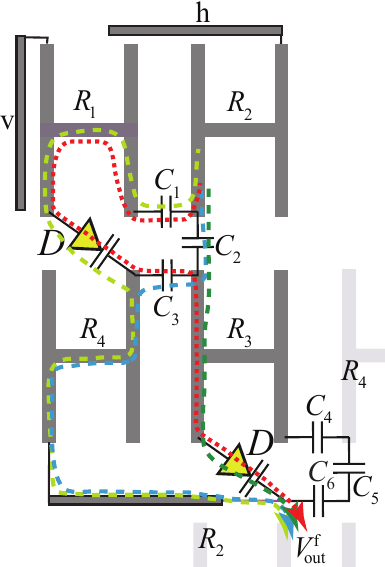}}
	\subfigure[]{\label{Fig:3c}
		\includegraphics[width=0.9\columnwidth]{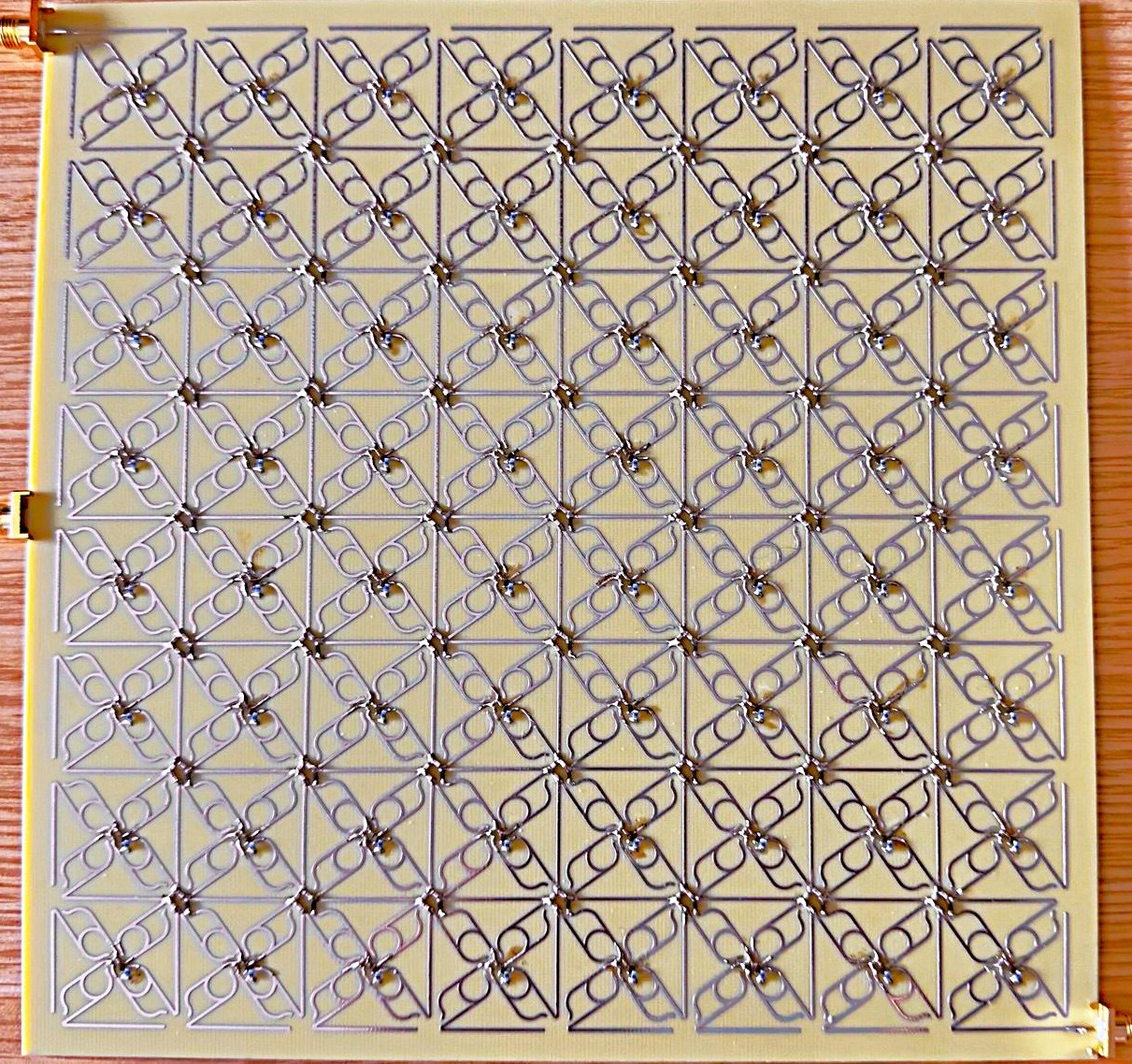}}	
	\caption{Unit-cell design and fabricated array. (a)~Unit-cell layout: horizontal and vertical dipoles double as free-space antennas and inter-cell interconnects, with the central varactor cluster ($L$, $D$, $C_1$--$C_3$) at the plaquette centre. (b)~Transmission-line model of the unit cell, showing sites $R_1$--$R_4$ and the Yellow, Green, Blue and Red propagation paths of \eqref{eq:pathdiff}. (c)~Photograph of the top conductor layer of the fabricated array. The array layout with TX and RX indicated and the bottom conductor layer are in the SM.}
	\label{fig:3}
\end{figure}

\emph{Synthetic gauge field and nonreciprocity.} Each unit cell carries two varactor clusters, both driven at $\Omega$ with independent phases $\phi_c$ and $\phi_{\rm br}$, $C_{c,\rm br}(t)=C_{c,\rm br\,0}+\Delta C_{c,\rm br}\cos(\Omega t+\phi_{c,\rm br})$ (Fig.~\ref{fig:3}). In the Floquet picture, a modulated bond couples sector $m$ to sectors $m\pm1$. Following the photonic Aharonov-Bohm effect~\cite{fang2012photonic}, the frequency-converting parts of the effective couplings for the two directions are a conjugate pair, while the static part $\kappa_0$ carries no phase,
\begin{equation}
\kappa_{ij}^{(\pm1)}=\frac{\delta\kappa}{2}\,e^{+i\phi},\qquad
\kappa_{ji}^{(\pm1)}=\frac{\delta\kappa}{2}\,e^{-i\phi}.
\label{eq:kappa}
\end{equation}
This is a Peierls phase $\phi$ on the frequency-converting channel of the bond, so closed paths that contain modulated bonds can enclose a nonzero synthetic flux, which breaks time-reversal symmetry. The bulk gap stays open in Regime~I.

For transmission, an input $V_{\rm in}$ at a dipole port excites four paths (Fig.~\ref{fig:3}b): Yellow (through the central $\pi$-flux bond), Green (through the bottom-right bond), Blue (static bonds only) and Red (through both modulated bonds). To first order in the modulation depths $m_c=\delta\gamma_c/\gamma$ and $m_{\rm br}=\delta\gamma_{\rm br}/\gamma$, the forward-minus-backward output voltage between the two ports of a unit cell is
\begin{equation}
\begin{split}
V^{\rm (f)}_{\rm out}-V^{\rm (b)}_{\rm out}=2iV_{\rm in}\big[&(t_{\rm Y}+t_{\rm R})m_c\sin\phi_c\\
&+(t_{\rm G}+t_{\rm R})m_{\rm br}\sin\phi_{\rm br}\big],
\end{split}
\label{eq:pathdiff}
\end{equation}
with $t_p$ the static transmission amplitude of path $p$ (derivation in the SM). This voltage difference vanishes for $\phi=0,\pi$ and is largest at $\phi_c=\phi_{\rm br}=90^\circ$, the operating point used throughout. The power contrast $|V^{\rm (f)}_{\rm out}|^2-|V^{\rm (b)}_{\rm out}|^2$ additionally depends on the relative phases of the static amplitudes $t_p$. To first order in $1/\Omega$, $\mathcal{H}_F=H_0+\sum_{n\neq0}H_nH_{-n}/(n\Omega)$~\cite{Goldman2015}, with $H_n$ the Fourier coefficient of $H(t)$ at $e^{in\Omega t}$. For $ f_\text{m}=2.346$~GHz the exact quasienergies of the Bloch Hamiltonian agree with those of the time-averaged $H_0$ to within $0.01$ in units of $2\pi\times$MHz (SM), so the band structure is that of the static lattice.

\emph{Sidebands and routing.} The modulated array generates sidebands at $f^{(m)}=f_0+m f_\text{m}$, $m=0,\pm1,\pm2,\dots$. A tone at $f_0$ injected at TX (top-left corner, $R_1$) is converted to $f_1=f_0+ f_\text{m}$ and radiated through the corner dipole. This is a resonance-matching effect: when $f_0$ lies far from the array resonance near $2.436$~GHz, which we associate with the corner modes, it is the $m=+1$ sideband that coincides with it and couples efficiently. A free-space wave at $f_1$ couples directly to this resonance and is delivered at the same frequency to RX (bottom-right corner, $R_3$). Conversion and routing are separate consequences of the mechanism, not a single chained process: the sideband is selected by resonance matching, whereas the direction in which captured energy is routed is set by the broken time-reversal symmetry, which predicts that the same wave is not returned to TX (Fig.~\ref{fig:1}b). We quantify the contrast between the two ports under identical illumination at $f_1$ with absolute powers read at the two ports. We use absolute power rather than $S$ parameters, because a swept network analyzer assumes a linear time-invariant device (SM).

\emph{Device and measurements.} The metasurface is a periodic array of four-resonator plaquettes on an FR4 substrate, fabricated by standard PCB processes (Fig.~\ref{fig:3}c). The intra-cell coupling $\gamma$ is set by fixed capacitors, the inter-cell couplings $\lambda_{\rm v}$ and $\lambda_{\rm h}$ by vertical and horizontal dipoles that double as antennas and interconnects and are not modulated. A central varactor cluster ($L$, $D_c$, $C_1$--$C_3$) and a bottom-right cluster ($D_{\rm br}$, $C_4$--$C_6$) provide the modulated bonds, each DC biased (from $1.5$ to $4.1$~V) and driven at $\Omega$. TX is the top-left corner ($R_1$) and RX the bottom-right corner ($R_3$). Two phase-locked signal generators supplied the RF tone at $f_0$ (TX, $-20$~dBm) and the modulation at $ f_\text{m}$ with controllable phases. The radiated field was recorded with a spectrum analyzer and a broadband horn antenna swept in angle in an anechoic chamber. For reception, a signal generator and a broadband antenna illuminated the array at $f_1$ with the modulation on, and the power at each port was read directly with the spectrum analyzer. The unmodulated array was first characterized with a vector network analyzer. Set-up details are in the SM.

\begin{figure}
	\begin{center}
\subfigure[]{\label{Fig:4a}
	\includegraphics[width=0.45\columnwidth]{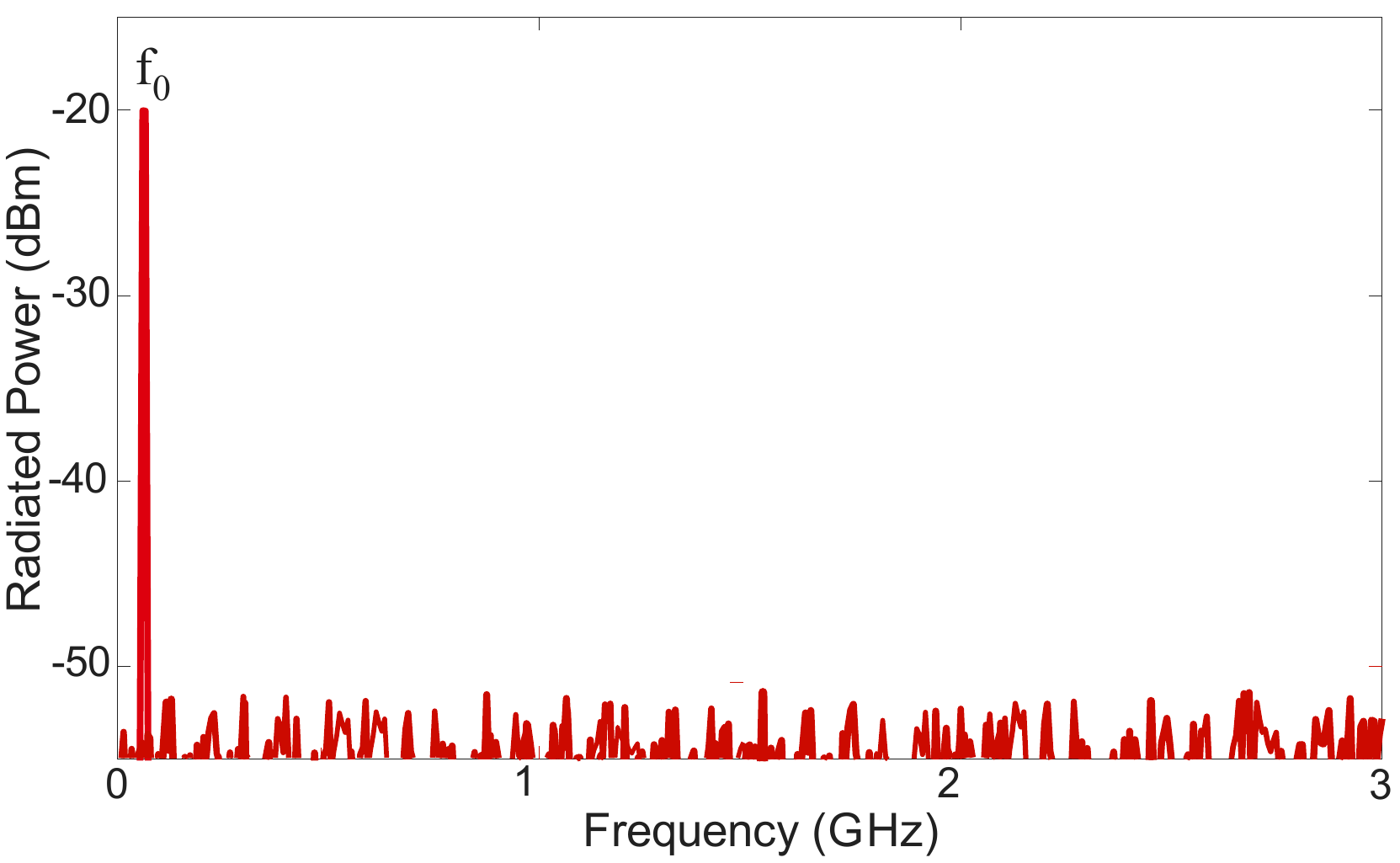}}
\subfigure[]{\label{Fig:4b}
	\includegraphics[width=0.45\columnwidth]{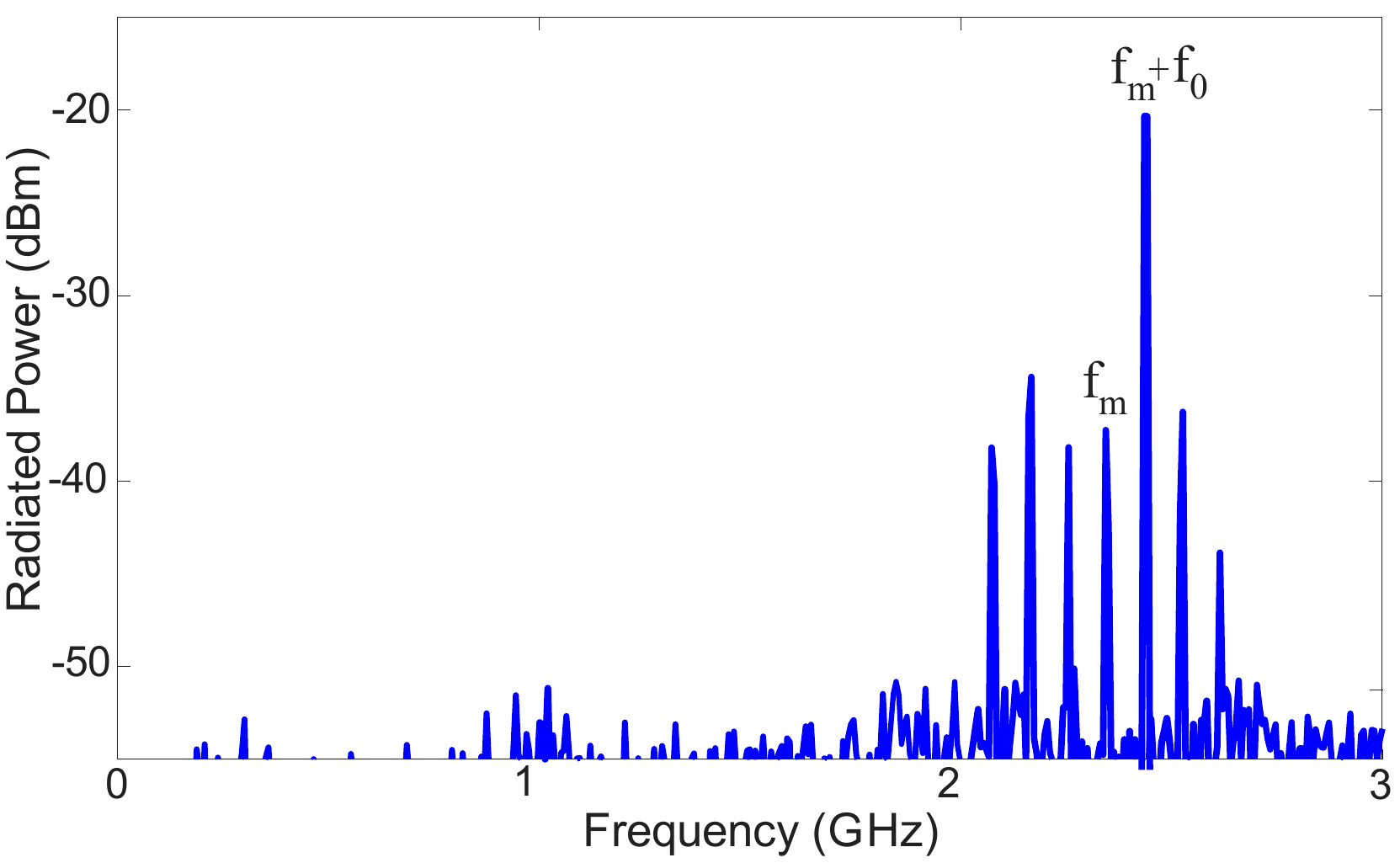}}
\subfigure[]{\label{Fig:4c}
	\includegraphics[width=0.6\columnwidth]{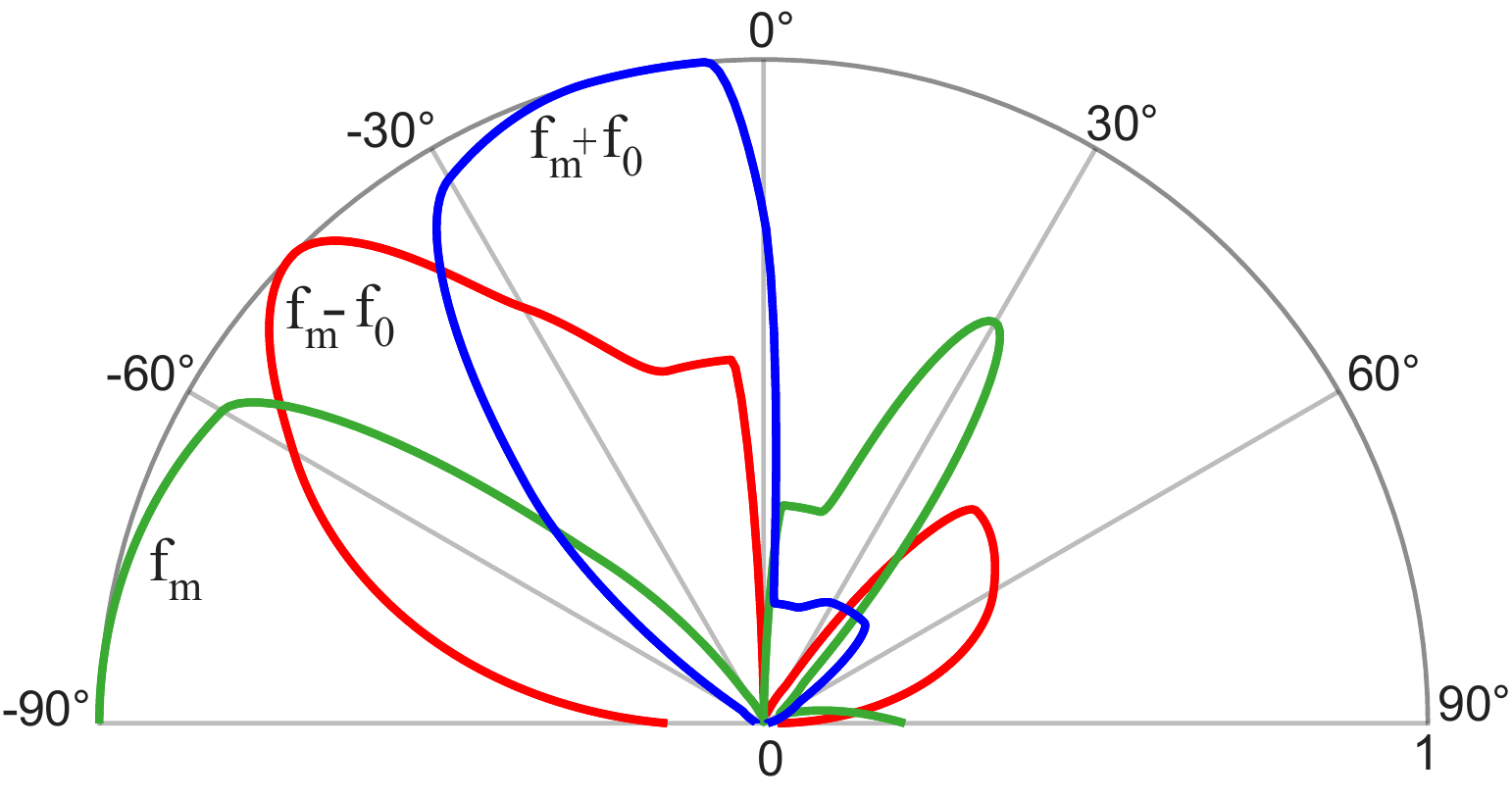}}	
		\caption{Frequency up-conversion and spatial decomposition of the time harmonics for $ f_\text{m}\gg f_0$, with $ f_\text{m}=2.346$~GHz and $f_0=90$~MHz. (a)~The TX port is excited at $f_0=90$~MHz. (b)~The radiated field is dominated by the up-converted harmonic at $f_{+1}= f_\text{m}+f_0=2.436$~GHz at broadside ($\theta_r=0^\circ$). (c)~Normalized radiation pattern in the $x$--$y$ plane, showing that the individual Floquet harmonics peak at distinct angles.}
		\label{fig:4}
	\end{center}
\end{figure}

\begin{figure*}
	\subfigure[]{\label{Fig:5a}
		\includegraphics[width=2\columnwidth]{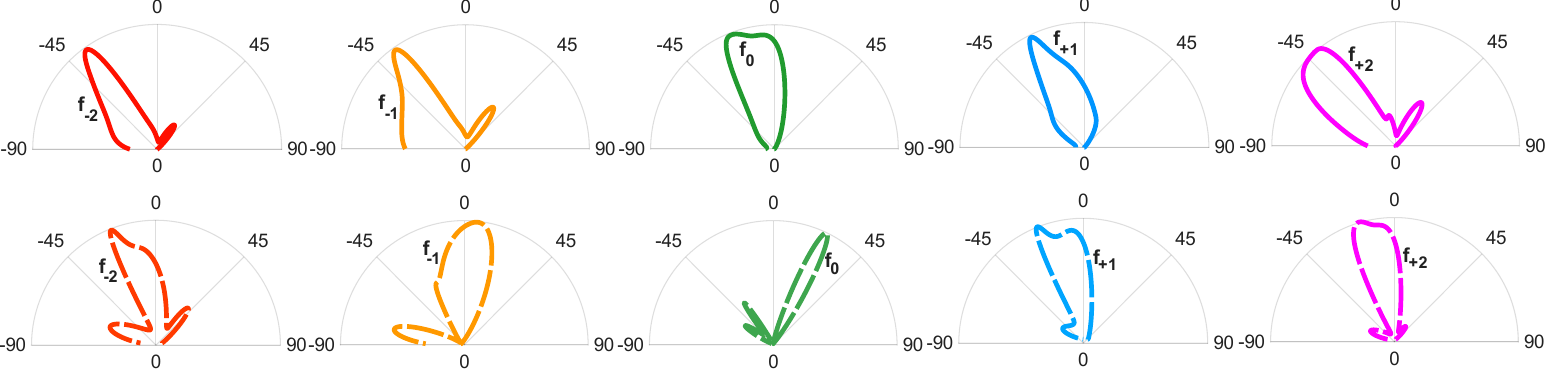}}
	\subfigure[]{\label{Fig:5b}
		\includegraphics[width=0.39\columnwidth]{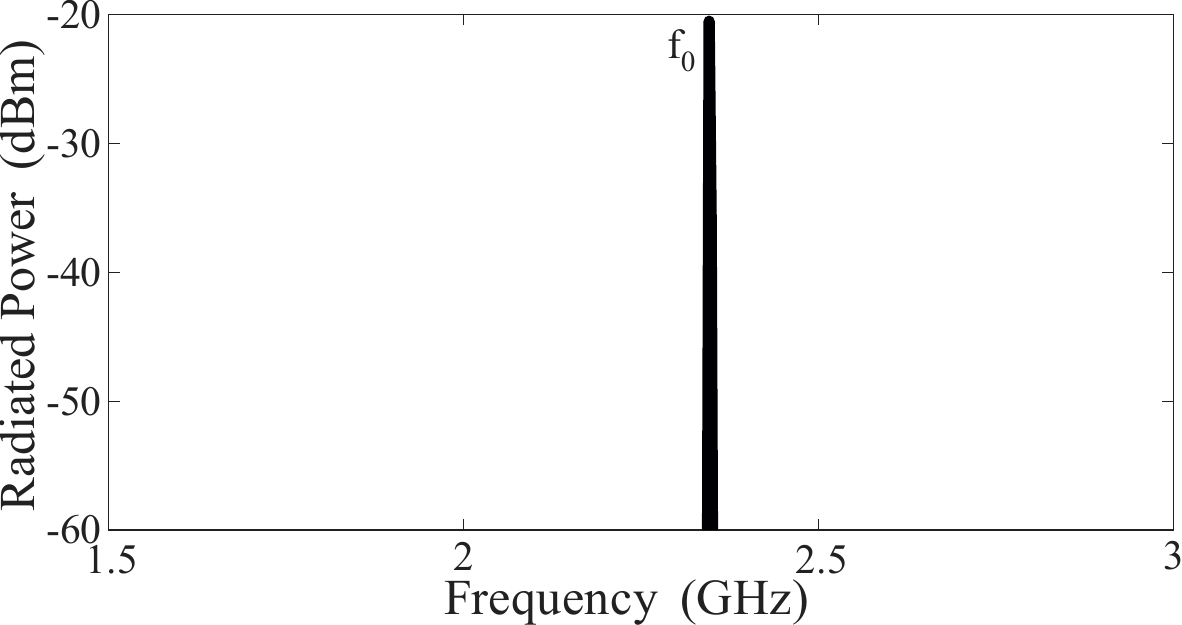}}
	\subfigure[]{\label{Fig:5c}
		\includegraphics[width=0.39\columnwidth]{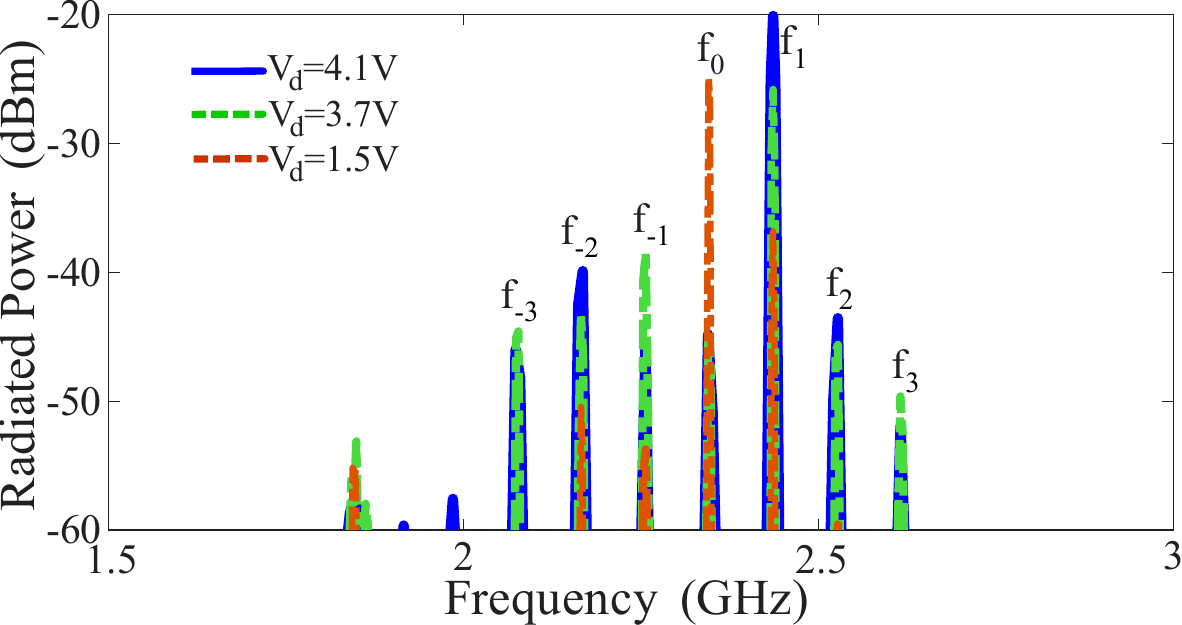}}	
	\subfigure[]{\label{Fig:5d}
		\includegraphics[width=0.39\columnwidth]{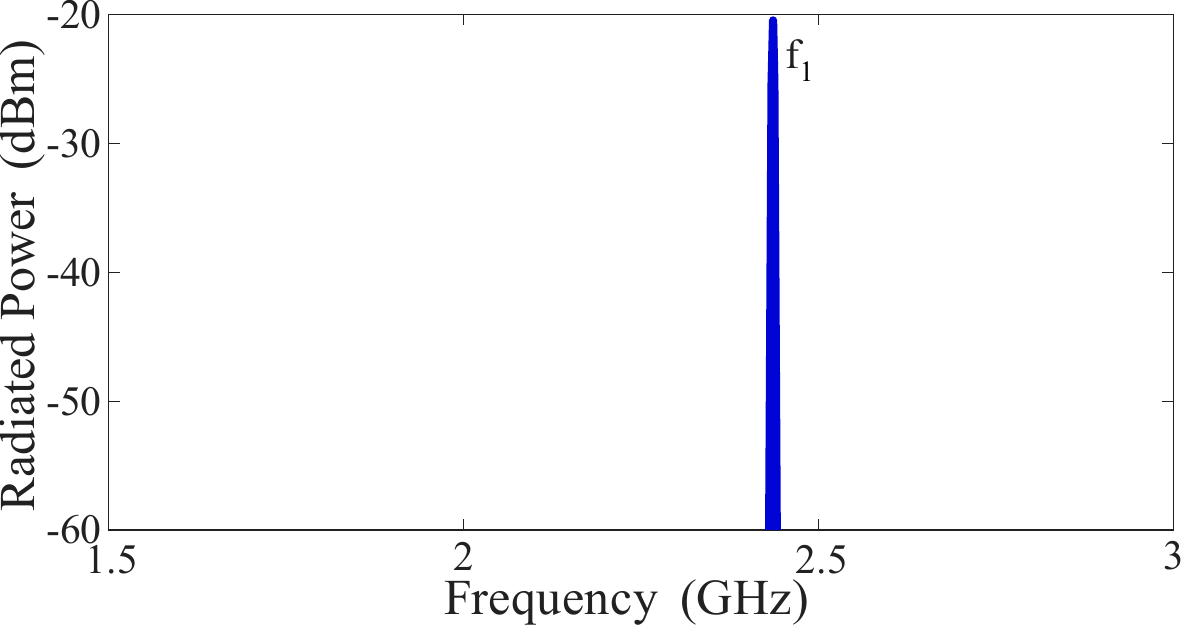}}
	\subfigure[]{\label{Fig:5e}
		\includegraphics[width=0.39\columnwidth]{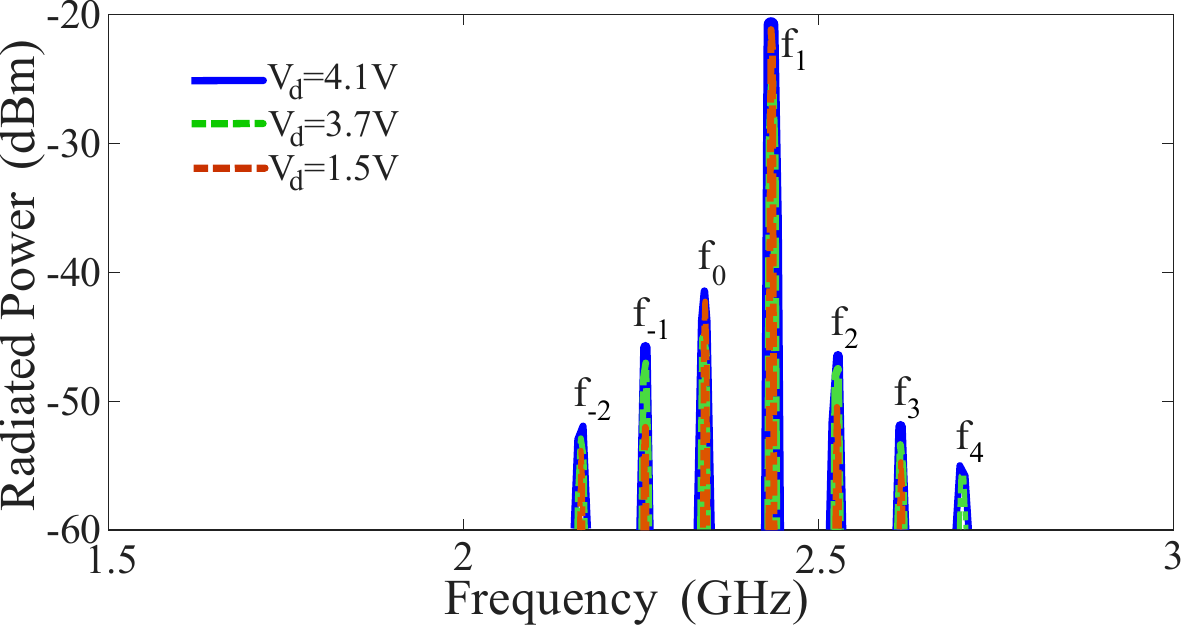}}
	\subfigure[]{\label{Fig:5f}
		\includegraphics[width=0.39\columnwidth]{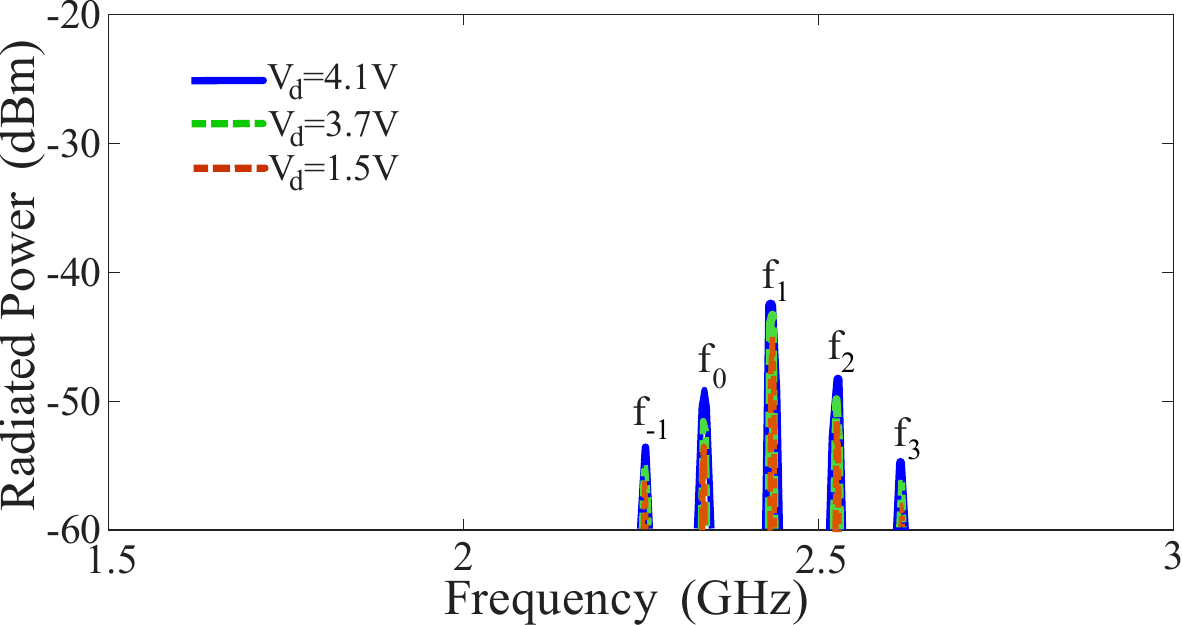}}		
	\caption{Frequency generation and nonreciprocal reception for $ f_\text{m}=90$~MHz and $f_0=2.436$~GHz; increasing the DC bias of the varactors from $1.5$ to $4.1$~V strengthens the harmonic generation. (a)~E-plane (top) and H-plane (bottom) radiation patterns at different sideband frequencies. (b)~TX-port excitation at $f_0$ leads to (c)~up-link frequency conversion from $f_0$ (TX port) to $f_1$ (air). (d)~The incoming wave from the air at $f_1$ shows no significant frequency generation at the source, and gives (e)~down-link reception at RX port at $f_1$ and (f)~weak reception at TX port.}
	\label{fig:5}
\end{figure*}

\emph{Experimental results.} We first tested up-conversion for $ f_\text{m}\gg f_0$, with $f_0=90$~MHz and $ f_\text{m}=2.346$~GHz. The injected tone is spectrally pure at $f_0$ (Fig.~\ref{fig:4}a), yet the radiated far field is dominated by the sideband $f_{+1}=f_0+ f_\text{m}=2.436$~GHz at broadside (Fig.~\ref{fig:4}b), consistent with the resonance-matching argument: $f_0$ lies far below the array resonance, and the $m=+1$ sideband coincides with it. The individual Floquet harmonics peak at distinct angles (Fig.~\ref{fig:4}c), so the angular and spectral separation of the comb persists under this large frequency translation.

We then tested the complementary regime $ f_\text{m}\ll f_0$, with $f_0=2.436$~GHz and $ f_\text{m}=90$~MHz. The TX excitation is again spectrally pure (Fig.~\ref{fig:5}b), but the radiated field resolves a Floquet comb spaced by $ f_\text{m}$ (Fig.~\ref{fig:5}c), each sideband has its own E-plane and H-plane pattern (Fig.~\ref{fig:5}a), and the sideband strength grows with the varactor bias from $1.5$ to $4.1$~V, confirming a modulation-driven effect rather than a fixed linear response. The two regimes, reached with modulation frequencies that differ by more than an order of magnitude, converge on the same radiated frequency, $\approx2.436$~GHz. We interpret this as a property of a resonance of the structure rather than of the choice of $f_0$ or $ f_\text{m}$.

Finally, we tested the reception leg. The array was illuminated from free space at $f_1=2.436$~GHz with a spectrally pure tone (Fig.~\ref{fig:5}d). At RX the tone is received dominantly at $f_1$ (Fig.~\ref{fig:5}e), with the other Floquet sidebands substantially weaker, while under the same illumination TX receives a markedly weaker signal (Fig.~\ref{fig:5}f). The power delivered to TX is more than $21$~dB below that delivered to RX, $\mathcal{I}=10\log_{10}[P_{\rm RX}(f_1)/P_{\rm TX}(f_1)]>21$~dB. This is the nonreciprocal reception expected from broken time-reversal symmetry: the incident wave is routed to RX at the frequency at which it arrived, and the path to TX is strongly suppressed.

\emph{Scope of the evidence and outlook.} The measurements establish modulation-driven frequency conversion together with nonreciprocal reception in an array built on the quadrupole lattice and operated in the gapped regime. The convergence of the two regimes on $\approx2.436$~GHz identifies a resonance of the structure, but by itself does not show that it is a topological corner mode. A spatial map of the corner-localized field and a modulation-off control of the port-power contrast are the most direct tests. Our Floquet calculation finds the zero-quasienergy corner modes but no anomalous $\pi$-quasienergy corner modes at these parameters (SM), so we make no claim about them. The first-order high-frequency expansion is controlled for $ f_\text{m}=2.346$~GHz but less clearly for $ f_\text{m}=90$~MHz, which is comparable to $\lambda_0$, where the quasienergy bands fold onto each other and we rely on the gap condition of \eqref{eq:regime} and on the numerical check. Extensions include higher multipoles, drives that produce anomalous $\pi$-quasienergy corner modes, Regime~II operation in which the gap closes during the cycle, a multi-port circulator using the two unused corners, and photonic, acoustic or superconducting-circuit platforms.

\newpage
\textbf{{\Large Supplemental Material}}

\section{Real-space tight-binding model}
The static quantized quadrupole insulator (the Benalcazar--Bernevig--Hughes, or BBH, model) maps onto a lattice of four-site plaquettes with intra-cell coupling $\gamma$ and inter-cell couplings $\lambda_{\rm v},\lambda_{\rm h}$. Labelling the four resonators of a unit cell $R_1$ (top-left), $R_2$ (top-right), $R_3$ (bottom-right), $R_4$ (bottom-left), the static tight-binding Hamiltonian is
\begin{equation}
	H_0 = \begin{pmatrix}
		0 & \gamma & \lambda_{\rm v} & -\gamma_{\rm D}^{+} \\
		\gamma & 0 & \gamma & \lambda_{\rm h} \\
		\lambda_{\rm v} & \gamma & 0 & \gamma-\lambda_{\rm hD}^{+} \\
		-\gamma_{\rm D}^{-} & \lambda_{\rm h} & \gamma-\lambda_{\rm hD}^{-} & 0
	\end{pmatrix},
	\label{eq:H0}
\end{equation}
where the negative sign on the $R_1$--$R_4$ bond implements a synthetic $\pi$ magnetic flux through the plaquette. This flux forces the two mirror symmetries of the lattice to anticommute, which is the algebraic origin of the quantized bulk quadrupole moment $q_{xy}=e/2$ and its associated corner charges $\pm e/2$~\cite{Benalcazar2017a,peterson2018quantized}.

The matrix elements of \eqref{eq:H0}, in the basis $(R_1,R_2,R_3,R_4)$ with identical bare resonance frequencies (zero diagonal in the rotating frame), are as follows. The coupling $\gamma$ is the intra-cell coupling between adjacent resonators of the same unit cell through fixed capacitors. The coupling $\lambda_{\rm v}$ is the vertical inter-cell coupling between $R_1$ of one cell and $R_3$ of the adjacent cell above or below, through the vertical dipole, which serves both as an antenna and as an inter-cell interconnect. The coupling $\lambda_{\rm h}$ is the horizontal inter-cell coupling between $R_2$ and $R_4$ of the adjacent cell to the left or right, through the horizontal dipole. The couplings $\gamma_{\rm D}^{\pm}$ are the intra-cell couplings between $R_1$ and $R_4$ through the anode-to-cathode ($+$) and cathode-to-anode ($-$) directions of the central varactor, with $\gamma_{\rm D}^{\pm}=\gamma$ in the static limit; the negative sign in \eqref{eq:H0} implements the $\pi$ flux. Likewise, $\lambda_{\rm hD}^{\pm}$ are the inter-cell couplings between $R_3$ and $R_2$ of adjacent cells through the two directions of the bottom-right varactor, mediated by the horizontal dipole.
\section{Bloch Hamiltonian and multipole relations}
Fourier transforming onto momentum space gives the Bloch Hamiltonian
\begin{equation}\label{eq:Hk}
	\begin{split}		
		H(\mathbf{k}) = &[\gamma+\lambda\cos k_x]\Gamma_4 + \lambda\sin k_x\,\Gamma_3 \\&+ [\gamma+\lambda\cos k_y]\Gamma_2 + \lambda\sin k_y\,\Gamma_1,
	\end{split}
\end{equation}
with $\Gamma_i$ the $4\times4$ gamma matrices of Ref.~\cite{Benalcazar2017a}; the corner-mode-supporting topological phase occurs for $|\gamma/\lambda|<1$. Here $\Gamma_k=-\tau_2\sigma_k$ for $k=1,2,3$ and $\Gamma_4=\tau_1\sigma_0$, with $\tau_i,\sigma_i$ Pauli matrices acting on the orbital (sublattice) degrees of freedom within a unit cell. These matrices encode the hopping pattern between the four sites of the unit cell that produces a $\pi$ flux through each plaquette, together with the mirror symmetries $M_x,M_y$ required for a quantized quadrupole phase.

The bulk quadrupole moment $q_{xy}$ is not simply an abstract invariant: it fixes the boundary observables directly. In the multipole framework introduced for the static quadrupole insulator~\cite{Benalcazar2017a}, the tangential edge polarizations $p_x,p_y$ and the corner charge $Q$ are related to the bulk moment by
\begin{equation}
	p_j^{\rm edge,\alpha} = n_i^\alpha q_{ij}, \qquad
	Q^{\rm corner,\alpha,\beta} = n_i^\alpha n_j^\beta q_{ij},
	\label{eq:multipole}
\end{equation}
where $\alpha,\beta=\pm x,\pm y$ label the edges and corners of the plaquette and $n_i^\alpha$ is the corresponding outward unit normal vector. In the topological phase of \eqref{eq:Hk}, all three boundary quantities share the same magnitude, $|p_x|=|p_y|=|Q|=|q_{xy}|=e/2$; crucially, the corner charge is an independent bulk-topological observable rather than simply the sum of the two edge polarizations that meet at it~\cite{Benalcazar2017a}.

\section{Floquet theory and invariants}
With $H(\mathbf{k},t)$ obtained from \eqref{eq:Hk} by $\gamma\to\gamma(t)$, $\lambda\to\lambda(t)$ (Eq.~(2) of the main text), the time-evolution operator and the effective Floquet Hamiltonian are as follows. This section is general background. For the drive of the main text, the numerical check in Sec.~\ref{sec:floquetcheck} finds only the zero-quasienergy corner modes.
Because $H(\mathbf{k},t+T)=H(\mathbf{k},t)$ with $T=2\pi/\Omega$, the system is governed by Floquet's theorem. Periodic driving is a powerful tool for engineering the quasienergy band structure by tuning the driving amplitude, frequency and phase~\cite{Goldman2015}, and the resulting topological classification is considerably richer than the static one: although quasienergy plays a role analogous to energy, in-gap boundary modes can be pinned at $\varepsilon=0$ and $\varepsilon=\pi/T$ even when the bulk quasienergy bands are themselves topologically trivial~\cite{Rudner2013}. The unitary time-evolution operator $U(t)=\mathcal{T}\exp\!\big[-\tfrac{i}{\hbar}\int_0^t H(\tau)\,d\tau\big]$ decomposes into a periodic micromotion (return-map) operator $\tilde U(t)=U(t)e^{iH_Ft}$ and the evolution generated by a time-independent effective Floquet Hamiltonian,
\begin{equation}
	H_F = \frac{i}{T}\log U(T), \qquad H_F\lvert u_n\rangle = \varepsilon_n\lvert u_n\rangle, \qquad \varepsilon_n\in\Big[-\frac{\pi}{T},\frac{\pi}{T}\Big],
	\label{eq:floquet}
\end{equation}
periodic under $\varepsilon\equiv\varepsilon+m\Omega$. The return map satisfies $\tilde U(0)=\tilde U(T)=I$ and captures the ``anomalous'' dynamical contribution to the topology that has no static analogue~\cite{Rudner2013}; it is this contribution that gives rise to the $\pi$-mode corner states discussed below. A periodically driven quadrupole insulator can support topological corner modes at \emph{two} distinct quasienergies~\cite{Rudner2013,Oka2019}: the ``0-mode'' at $\varepsilon=0$, the adiabatic continuation of the static BBH corner state and captured by $H_F$ together with the nested Wilson-loop invariant~\cite{Benalcazar2017b}; and the ``$\pi$-mode'' at $\varepsilon=\pi/T$, a purely dynamical state arising from the return map $\tilde U(t)$ with no static analogue, exhibiting period doubling with period $2T$. The two families are classified by independent $\mathbb{Z}_2$ invariants,
\begin{equation}
	\nu_0 = (n_0+\nu_0^F)\bmod 2, \qquad \nu_\pi = n_\pi \bmod 2,
	\label{eq:invariants}
\end{equation}
where $\nu_0^F$ is the static-like quadrupole invariant of $H_F$ and $n_0,n_\pi$ are dynamical invariants of $\tilde U(t)$~\cite{Peng2019,Huang2020}.

The static-like invariant $\nu_0^F$ follows from the nested polarization formula derived from the Wannier bands~\cite{Benalcazar2017b},
\begin{equation}
	p_y^j = i\int_{\rm BZ}\frac{d^2k}{(2\pi)^2}\,\langle\alpha_{x,k}^j\rvert\partial_{k_y}\lvert\alpha_{x,k}^j\rangle,
	\label{eq:nestedpol}
\end{equation}
where $\lvert\alpha_{x,k}^j\rangle$ are eigenstates of the Wannier Hamiltonian. In the presence of the mirror symmetries, these nested polarizations are quantized to $0$ (trivial) or $1/2$ (topological), and the topological quadrupole phase corresponds to
\begin{equation}
	(p_y^j,p_x^j)=(1/2,1/2).
	\label{eq:quadphase}
\end{equation}
\section{Gap-closing condition and operating regime}
The quadrupole phase is lost at the instants when the instantaneous bulk gap closes, i.e., when $\left\lvert \gamma(t)/\lambda(t) \right\rvert = 1$. Substituting Eqs.~(2a) and (2b) of the main text,
\begin{equation}
	\lvert\gamma_0+\delta\gamma\cos(\Omega t)\rvert = \lvert\lambda_0+\delta\lambda\sin(\Omega t)\rvert.
	\label{eq:transitioneq}
\end{equation}
Assuming the modulation is small enough that both arguments remain positive ($\gamma_0>\delta\gamma$, $\lambda_0>\delta\lambda$), the absolute values can be dropped,
\begin{align}
	\gamma_0+\delta\gamma\cos(\Omega t) &= \lambda_0+\delta\lambda\sin(\Omega t) \nonumber\\
	\Longrightarrow\;\; \delta\gamma\cos(\Omega t)-\delta\lambda\sin(\Omega t) &= \lambda_0-\gamma_0.
	\label{eq:rearranged}
\end{align}
This has the standard form $A\cos x+B\sin x=C$ with $A=\delta\gamma$, $B=-\delta\lambda$ and $C=\lambda_0-\gamma_0$, whose solution is
\begin{align}
	\Omega t = &-\arctan\!\left(\frac{\delta\lambda}{\delta\gamma}\right) \nonumber\\
	&\pm \arccos\!\left(\frac{\lambda_0-\gamma_0}{\sqrt{\delta\gamma^2+\delta\lambda^2}}\right) + 2\pi n, \quad n\in\mathbb{Z}.
	\label{eq:transitiontimes}
\end{align}
Real solutions of \eqref{eq:transitiontimes} exist only if the arccos argument does not exceed unity, which separates two regimes. In Regime~I, $\lambda_0-\gamma_0>\sqrt{\delta\gamma^2+\delta\lambda^2}$, no real solution exists and the system remains topological throughout the modulation cycle. In Regime~II, $\lambda_0-\gamma_0<\sqrt{\delta\gamma^2+\delta\lambda^2}$, the system crosses the phase transition at the times given by \eqref{eq:transitiontimes}. For the fabricated array, $\gamma_0=35$~MHz and $\lambda_0=150$~MHz (static ratio $\lambda_0/\gamma_0\approx4.3$), with modulation amplitudes $\delta\gamma\approx10$~MHz and $\delta\lambda\approx20$~MHz. Then $\lambda_0-\gamma_0=115$~MHz exceeds $\sqrt{\delta\gamma^2+\delta\lambda^2}\approx22.4$~MHz, so the device operates in Regime~I and never crosses the phase transition.
\section{Synthetic gauge field and modulation matrix}
A quantized quadrupole phase cannot be reached by simply coupling one-dimensional SSH chains: it requires a gauge field that opens a bulk gap and modifies the mirror-symmetry algebra. This is achieved by a synthetic magnetic flux of $\pi$ through each plaquette, which (i) opens a gap at half filling and (ii) changes the commutation relation of the two mirror symmetries from commuting to anticommuting, the algebraic foundation of the quantized quadrupole moment and of the fractional corner charges~\cite{Benalcazar2017a,peterson2018quantized}. In the same way that a charged particle encircling a real magnetic flux acquires the Aharonov--Bohm phase $\Phi_B=\oint\mathbf{A}\cdot d\mathbf{l}$, a wave circulating a plaquette acquires a phase set by the synthetic gauge field, although no real magnetic field is present. Physically, two resonators are positively coupled ($+\lambda$ or $+\gamma$) when their oscillations are in phase, which is realized by connecting the coupling capacitor to anti-nodes of the same polarity. A negative coupling ($-\lambda$ or $-\gamma$) arises when the resonators are connected through anti-nodes of opposite polarity, which introduces an effective $\pi$ phase shift. The quadrupole phase requires exactly one negative coupling per plaquette, the other three being positive, so that paths circulating the plaquette interfere destructively, which realizes the $\pi$-flux gauge field.

The two varactors per unit cell are driven at the common frequency $\Omega$ but with independent phases $\phi_c,\phi_{\rm br}$.
The central varactor cluster (inductor $L$, varactor $D_c$ and fixed capacitors $C_1$--$C_3$) is connected across the $R_1$--$R_4$ bond and modulates both $\gamma_{\rm D}^{+}$ and $\gamma_{\rm D}^{-}$, while the bottom-right varactor cluster (varactor $D_{\rm br}$ and fixed capacitors $C_4$--$C_6$) is connected between $R_3$ and $R_4$ and influences the inter-cell couplings $\lambda_{\rm hD}^{+}$ and $\lambda_{\rm hD}^{-}$. The time-varying capacitances are
\begin{subequations}
	\begin{align}
		C_c(t) &= C_{c0}+\Delta C_c\cos(\Omega t+\phi_c),\\
		C_{\rm br}(t) &= C_{\rm br0}+\Delta C_{\rm br}\cos(\Omega t+\phi_{\rm br}).
	\end{align}
	\label{eq:capacitances}
\end{subequations}

In the Floquet picture a modulated bond couples sector $m$ to sectors $m\pm1$. Following the photonic Aharonov--Bohm effect~\cite{fang2012photonic,taravati2020full,taravati2022low}, the frequency-converting parts of the effective couplings for forward and backward propagation are a conjugate pair, while the static part $\kappa_0$ carries no phase,
\begin{equation}
	\kappa_{ij}^{(\pm1)}=\frac{\delta\kappa}{2}\,e^{+i\phi},\qquad
	\kappa_{ji}^{(\pm1)}=\frac{\delta\kappa}{2}\,e^{-i\phi}.
	\label{eq:asymcoupling}
\end{equation}
This is a Peierls phase $\phi$ on the frequency-converting channel of the bond, and it is the source of the direction-dependent coupling in the device, as captured by the entries of $V(t)$ below.

Collecting the modulation-induced changes in a perturbation matrix $V(t)$, the real-space tight-binding Hamiltonian of a single unit cell is $H(t)=H_0+V(t)$, with
\begin{equation}
	V(t) = \begin{pmatrix}
		0 & 0 & 0 & \delta\gamma_{14}(t) \\
		0 & 0 & -\delta\lambda_{23}(t) & \delta\lambda_{24}(t) \\
		0 & -\delta\lambda_{32}(t) & 0 & -\delta\gamma_{34}(t) \\
		-\delta\gamma_{41}(t) & -\delta\lambda_{42}(t) & -\delta\gamma_{43}(t) & 0
	\end{pmatrix},
	\label{eq:Vt}
\end{equation}
and direction-dependent entries
\begin{equation}
	\begin{aligned}
		\delta\gamma_{14}(t) &= \delta\gamma_c\, e^{+i\phi_c}\cos(\Omega t),\\
		\delta\gamma_{41}(t) &= \delta\gamma_c\, e^{-i\phi_c}\cos(\Omega t),\\
		\delta\gamma_{34}(t) &= \delta\gamma_{\rm br}\, e^{+i\phi_{\rm br}}\cos(\Omega t),\\
		\delta\gamma_{43}(t) &= \delta\gamma_{\rm br}\, e^{-i\phi_{\rm br}}\cos(\Omega t),\\
		\delta\lambda_{32}(t) &= \delta\lambda_{\rm br}\, e^{+i\phi_{\rm br}}\cos(\Omega t),\\
		\delta\lambda_{42}(t) &= \delta\lambda_{\rm br}\, e^{-i\phi_{\rm br}}\cos(\Omega t),\\
		\delta\lambda_{23}(t) &= \delta\lambda_{\rm br}\, e^{+i\phi_{\rm br}}\cos(\Omega t),\\
		\delta\lambda_{24}(t) &= \delta\lambda_{\rm br}\, e^{-i\phi_{\rm br}}\cos(\Omega t),
	\end{aligned}
	\label{eq:dentries}
\end{equation}
so that, for each modulated bond, the two directions carry conjugate phases, e.g., $\delta\gamma_{14}(t)=e^{+2i\phi_c}\,\delta\gamma_{41}(t)$.
\section{Path interference and nonreciprocal transmission}
Wave propagation inside the unit cell is described by voltage phasors $\tilde V(\mathbf{r},t)=\mathrm{Re}[\tilde V(\mathbf{r})e^{-i\omega t}]$ at the anti-nodal points. An input voltage $V_{\rm in}$ at a dipole port excites several propagation paths (Yellow, Green, Blue and Red; Fig.~2(b) of the main text), and the output voltage is their coherent superposition,
\begin{equation}
	\tilde V_{\rm out}(\omega) = \sum_p t_p(\omega)\,M_p(\phi_c,\phi_{\rm br})\,V_{\rm in},
	\label{eq:superposition}
\end{equation}
where $t_p(\omega)$ is the static transmission amplitude of path $p$ and $M_p$ is its Floquet modulation factor. The Yellow path traverses the $R_1$--$R_4$ bond (the $\pi$-flux bond), modulated by the central varactor $D_c$: the forward direction uses $\gamma_{\rm D}^{+}(t)$ with phase $+\phi_c$ and the backward direction uses $\gamma_{\rm D}^{-}(t)$ with phase $-\phi_c$. The Green path traverses the $R_3$--$R_4$ bond, modulated by the bottom-right varactor $D_{\rm br}$, with phases $\pm\phi_{\rm br}$ in the forward and backward directions. The Blue path traverses only static bonds, such as $R_1$--$R_2$ or $R_2$--$R_3$ through $\gamma$, and carries no modulation. The Red path traverses both modulated bonds in sequence, and to first order in the modulation depths the product of the individual factors reduces to a sum. With $m_c=\delta\gamma_c/\gamma$ and $m_{\rm br}=\delta\gamma_{\rm br}/\gamma$, the modulation factors are
\begin{equation}
	\begin{aligned}
		M_{\rm Y}^{\rm (f)} &= 1+m_c\,e^{+i\phi_c}, \qquad M_{\rm Y}^{\rm (b)} = 1+m_c\,e^{-i\phi_c},\\
		M_{\rm G}^{\rm (f)} &= 1+m_{\rm br}\,e^{+i\phi_{\rm br}}, \qquad M_{\rm G}^{\rm (b)} = 1+m_{\rm br}\,e^{-i\phi_{\rm br}},\\
		M_{\rm B}^{\rm (f)} &= M_{\rm B}^{\rm (b)} = 1,\\
		M_{\rm R}^{\rm (f)} &= 1+m_c\,e^{+i\phi_c}+m_{\rm br}\,e^{+i\phi_{\rm br}},\\
		M_{\rm R}^{\rm (b)} &= 1+m_c\,e^{-i\phi_c}+m_{\rm br}\,e^{-i\phi_{\rm br}}.
	\end{aligned}
	\label{eq:Mfactors}
\end{equation}
The total forward and backward output voltages are the coherent sums over all paths,
\begin{align}
	V_{\rm out}^{\rm (f)} = V_{\rm in}\Big[&t_{\rm B} + t_{\rm Y}\big(1+m_c e^{+i\phi_c}\big) \nonumber\\
	&+ t_{\rm G}\big(1+m_{\rm br}e^{+i\phi_{\rm br}}\big) \nonumber\\
	&+ t_{\rm R}\big(1+m_c e^{+i\phi_c}+m_{\rm br}e^{+i\phi_{\rm br}}\big)\Big], \label{eq:Vf}\\
	V_{\rm out}^{\rm (b)} = V_{\rm in}\Big[&t_{\rm B} + t_{\rm Y}\big(1+m_c e^{-i\phi_c}\big) \nonumber\\
	&+ t_{\rm G}\big(1+m_{\rm br}e^{-i\phi_{\rm br}}\big) \nonumber\\
	&+ t_{\rm R}\big(1+m_c e^{-i\phi_c}+m_{\rm br}e^{-i\phi_{\rm br}}\big)\Big], \label{eq:Vb}
\end{align}
where $t_{\rm B},t_{\rm Y},t_{\rm G},t_{\rm R}$ are the static transmission coefficients of the Blue, Yellow, Green and Red paths. The forward--backward difference is
\begin{align}
	V_{\rm out}^{\rm (f)}-V_{\rm out}^{\rm (b)} = V_{\rm in}\Big[&(t_{\rm Y}+t_{\rm R})\,m_c\big(e^{+i\phi_c}-e^{-i\phi_c}\big) \nonumber\\
	&+ (t_{\rm G}+t_{\rm R})\,m_{\rm br}\big(e^{+i\phi_{\rm br}}-e^{-i\phi_{\rm br}}\big)\Big].
	\label{eq:diff1}
\end{align}
Using $e^{+i\theta}-e^{-i\theta}=2i\sin\theta$, the forward\,--\,backward output-voltage difference becomes

\begin{equation}
	V_{\rm out}^{\rm (f)}-V_{\rm out}^{\rm (b)} = 2iV_{\rm in}\Big[(t_{\rm Y}+t_{\rm R})\,m_c\sin\phi_c + (t_{\rm G}+t_{\rm R})\,m_{\rm br}\sin\phi_{\rm br}\Big],
	\label{eq:pathdiff}
\end{equation}
where $m_c=\delta\gamma_c/\gamma$, $m_{\rm br}=\delta\gamma_{\rm br}/\gamma$; this voltage difference is largest at $\phi_c=\phi_{\rm br}=90^\circ$, the operating point used throughout. Forward and backward refer to the two directions between the two dipole ports of a unit cell. The power contrast $|V_{\rm out}^{\rm (f)}|^2-|V_{\rm out}^{\rm (b)}|^2$ additionally depends on the relative phases of the static amplitudes $t_p$. Crucially, this asymmetry does not close the bulk gap as long as the time-averaged coupling ratio remains topological, $\langle\lambda(t)\rangle/\langle\gamma(t)\rangle=\lambda_0/\gamma_0>1$; the gap therefore stays open while the transmission becomes direction-dependent.

The Floquet Hamiltonian in the high-frequency expansion to first order in $1/\Omega$~\cite{Goldman2015},
\begin{equation}
	\mathcal{H}_F = H_0+\sum_{n\neq0}\frac{H_nH_{-n}}{n\Omega},
	\label{eq:HFexp}
\end{equation}
where $H_n$ is the coefficient of $e^{in\Omega t}$ in the Fourier series of $H(t)$ (with $\hbar=1$), inherits the band topology of $H_0$ because the time modulation does not close the bulk band gap when $\lambda/\gamma>1$. Consequently, at high modulation frequency the quasienergy band structure is that of the static lattice (Sec.~\ref{sec:floquetcheck}). The expansion is less clearly controlled for $f_\text{m}=90$~MHz, which is comparable to $\lambda_0$. The mechanism therefore combines three ingredients: (i) a static $\pi$ flux, encoded in the negative sign of the $R_1$--$R_4$ coupling in $H_0$, which opens the band gap and protects the corner modes; (ii) asymmetric effective couplings arising from the photonic Aharonov--Bohm effect, where forward propagation acquires $e^{+i\phi}$ and backward propagation $e^{-i\phi}$ for each time-modulated bond, as captured in $\gamma_{\rm D}^{\pm}$, $\lambda_{\rm hD}^{\pm}$ and $V(t)$; and (iii) multiple modulated paths (Yellow, Green, Red) that sum coherently, with nonreciprocal contributions that add constructively at $\phi_c=\phi_{\rm br}=90^\circ$.

The same nonreciprocity can be viewed as direction-dependent interference between the propagation paths. The phase accumulated along path $m\in\{\text{Yellow},\text{Blue},\text{Red},\text{Green}\}$ in direction $s=\pm$ (clockwise or counter-clockwise around the array) is
\begin{equation}
	\Phi_m^{s}(t)=\int_0^t \omega_m^{s}(\tau)\,d\tau,
	\label{eq:pathphase}
\end{equation}
where $\omega_m^{\pm}$ is the instantaneous path frequency, which picks up opposite modulation phases $\pm\phi$ in the two directions (\eqref{eq:asymcoupling}). The accumulated phase sums for clockwise (CW) and counter-clockwise (CCW) propagation are
\begin{equation}
	\Delta\Phi_{\rm CW}=\sum_m\Phi_m^{+}, \qquad \Delta\Phi_{\rm CCW}=\sum_m\Phi_m^{-},
	\label{eq:DeltaPhi}
\end{equation}
which are generally unequal owing to the time modulation, producing direction-dependent interference. The forward and backward transmission may then be written as
\begin{align}
	S_{21}&\propto\left\lvert A\,e^{i\Delta\Phi_{\rm CW}}+B\,e^{i\Delta\Phi_{\rm interfere}}\right\rvert^2, \nonumber\\
	S_{12}&\propto\left\lvert A\,e^{i\Delta\Phi_{\rm CCW}}+B\,e^{i\Delta\Phi_{\rm interfere}}\right\rvert^2,
	\label{eq:S21S12}
\end{align}
where $A$ and $B$ are amplitude factors and $\Delta\Phi_{\rm interfere}$ is the direction-independent phase of the second interfering contribution. The resulting asymmetry $S_{21}\neq S_{12}$ is the signature of nonreciprocal propagation, in analogy with synthetic-gauge-field-induced directional transport in anomalous Floquet topological systems and non-Hermitian circuits.
\section{Floquet sidebands, up-conversion and reception efficiencies}
Because the corner resonance is periodically driven, it does not merely oscillate at a fixed frequency: it generates Floquet sidebands at
\begin{equation}
	f_{\rm corner}^{(m)} = f_0+mf_\text{m}, \qquad m=0,\pm1,\pm2,\dots.
	\label{eq:harmonics}
\end{equation}
A signal at $f_0$ launched into TX port (the top-left corner, site $R_1$) excites the corner mode, which is up-converted by the modulation and radiated into free space at $f_1=f_0+f_\text{m}$ through the corner's embedded dipole antenna (illustrated in Fig.~1(b) of the main text), with conversion efficiency
\begin{equation}
	\eta_{\rm up} = \frac{P_{\rm rad}(f_0+f_\text{m})}{P_{\rm in}(f_0)}.
	\label{eq:etaup}
\end{equation}
The radiated wave is captured in the far field by a broadband receive antenna. This up-conversion is fundamentally a resonance-matching effect: the injected tone at $f_0$ sits far below the array's own high-$Q$ operating frequency, whereas the $m=+1$ sideband coincides with it, so it is the sideband, and not the bare fundamental, that couples efficiently to the corner state and radiates.

Frequency conversion and directional routing are two separate consequences of the corner-state mechanism, not a single chained process. Illuminating the array with a free-space wave at $f_1$ couples directly, on resonance, into the corner mode, and is delivered at RX port (the bottom-right corner, site $R_3$) at the \emph{same} frequency $f_1$, with reception efficiency
\begin{equation}
	\eta_{\rm RX} = \frac{P_{2}(f_1)}{P_{\rm inc}(f_1)}.
	\label{eq:etaRX}
\end{equation}
Crucially, this reception is nonreciprocal: because time-reversal symmetry is broken by the conjugate couplings of \eqref{eq:asymcoupling}, the same incident wave at $f_1$ is not returned to TX port. \eqref{eq:etaup} therefore governs \emph{which} Floquet sideband couples efficiently to the corner mode, while the broken time-reversal symmetry governs \emph{where} the resulting energy is subsequently routed; the two need not, and in our device do not, involve a second frequency conversion. The device implements a TX port-$\to$-air-$\to$-RX port circulator: energy launched at TX port is converted in frequency and radiated, and any of that energy recaptured by the array is preferentially delivered at RX port, unconverted, while the reverse itinerary (RX port to air to TX port) is isolated. The isolation between the two directions is quantified by the standard transmission asymmetry evaluated at the Floquet sideband frequencies $f=f_0+mf_\text{m}$, though, as discussed in Sec.~\ref{sec:meas}, we verify it directly using absolute-power measurements rather than a swept $S$-parameter measurement, since the latter is not generally reliable for a polychromatic, time-modulated network.

A design in which the receiver instead down-converts the incident wave back to $f_0$ would replace $\eta_{\rm RX}$ by
\begin{equation}
	\eta_{\rm down}=\frac{P_{\rm RX}(f_0)}{P_{\rm inc}(f_0+f_\text{m})},
	\label{eq:etadown}
\end{equation}
with total transceiver efficiency $\eta_{\rm total}=\eta_{\rm up}\,\eta_{\rm down}$. This is possible in principle but is not the configuration realized in the present device.
\section{Device, measurement set-up and the case against $S$-parameters}\label{sec:meas}
\begin{figure}
	\includegraphics[width=1\columnwidth]{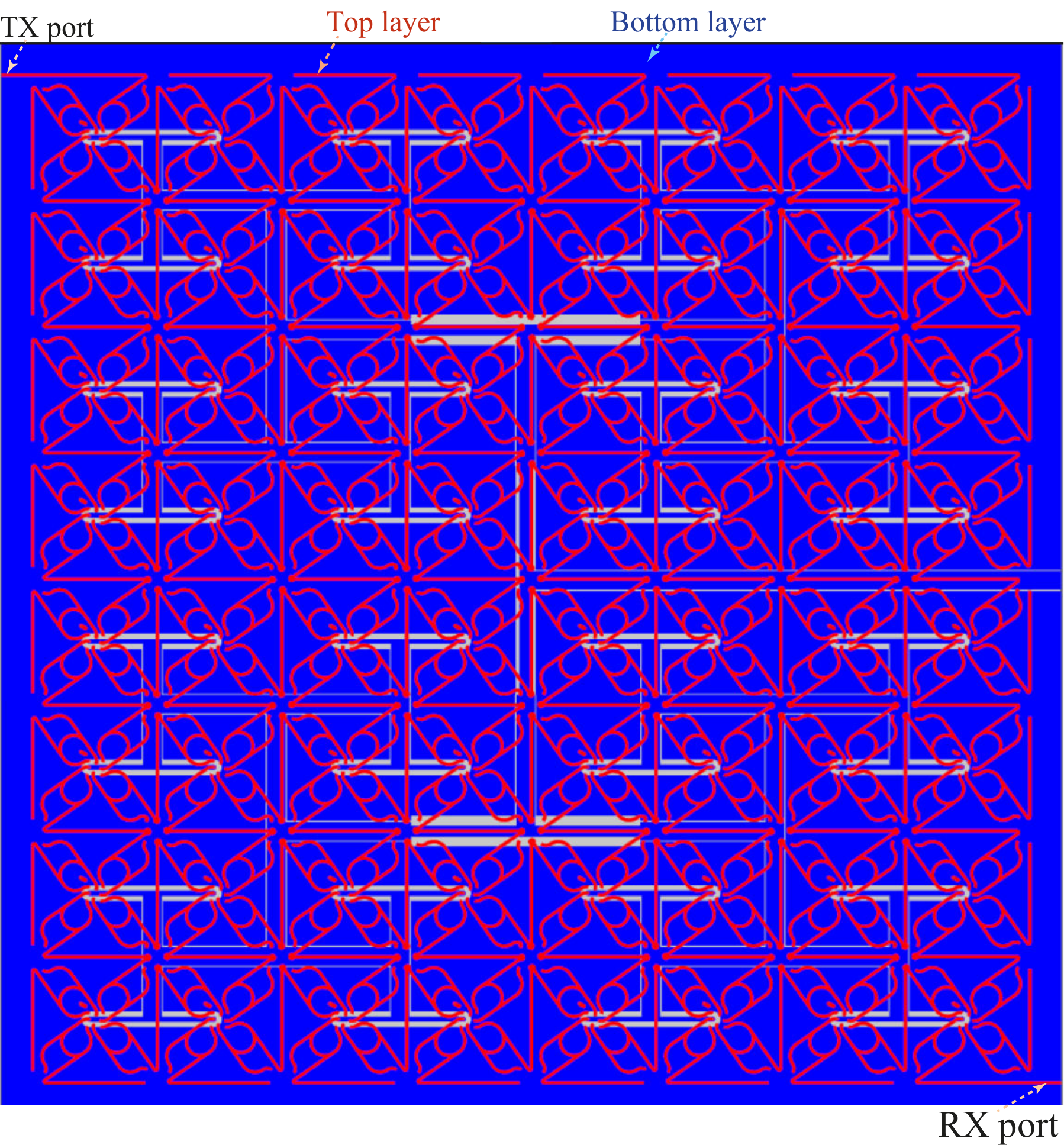}\hfill
	\includegraphics[width=1\columnwidth]{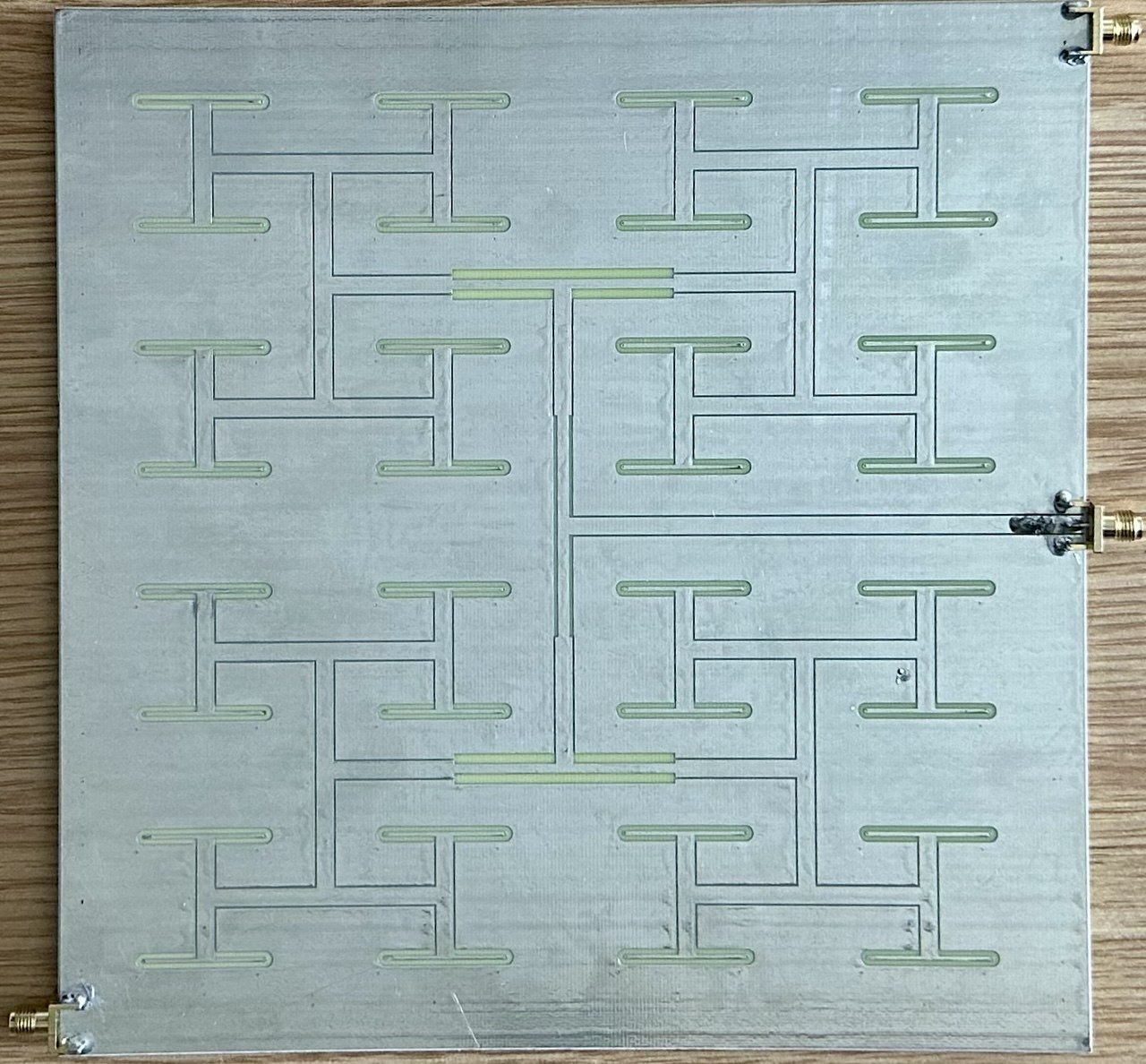}
	\caption{Fabricated array. (a)~Top conductor layer of the array, with TX port (top-left) and RX port (bottom-right) indicated. (b)~Bottom conductor layer of the fabricated prototype. The unit-cell layout and a photograph of the top layer are in Fig.~2 of the main text.}
	\label{Fig:Resba}
\end{figure}

Each unit cell realizes the four-site quadrupole plaquette of \eqref{eq:H0} with intra-cell coupling $\gamma$ (fixed capacitors) and inter-cell couplings $\lambda_{\rm v},\lambda_{\rm h}$ mediated by vertical and horizontal microstrip dipoles that double as free-space antennas and inter-cell interconnects. A central varactor cluster ($\gamma_{\rm D}^{\pm}$) and a bottom-right varactor cluster ($\lambda_{\rm hD}^{\pm}$) are independently biased and modulated at $\Omega$ with phases $\phi_c$ and $\phi_{\rm br}$, imprinting the direction-dependent couplings of \eqref{eq:asymcoupling}. The full tight-binding Hamiltonian, capacitance-modulation waveforms, and path-modulation factors are given in \eqref{eq:H0}, \eqref{eq:capacitances} and \eqref{eq:Mfactors}. The array was fabricated on a FR4 microwave substrate using standard PCB processes (Fig.~2 of the main text and Fig.~\ref{Fig:Resba}: top and bottom conductor layers, with TX port and RX port indicated).

\begin{figure}
	\includegraphics[width=1\columnwidth]{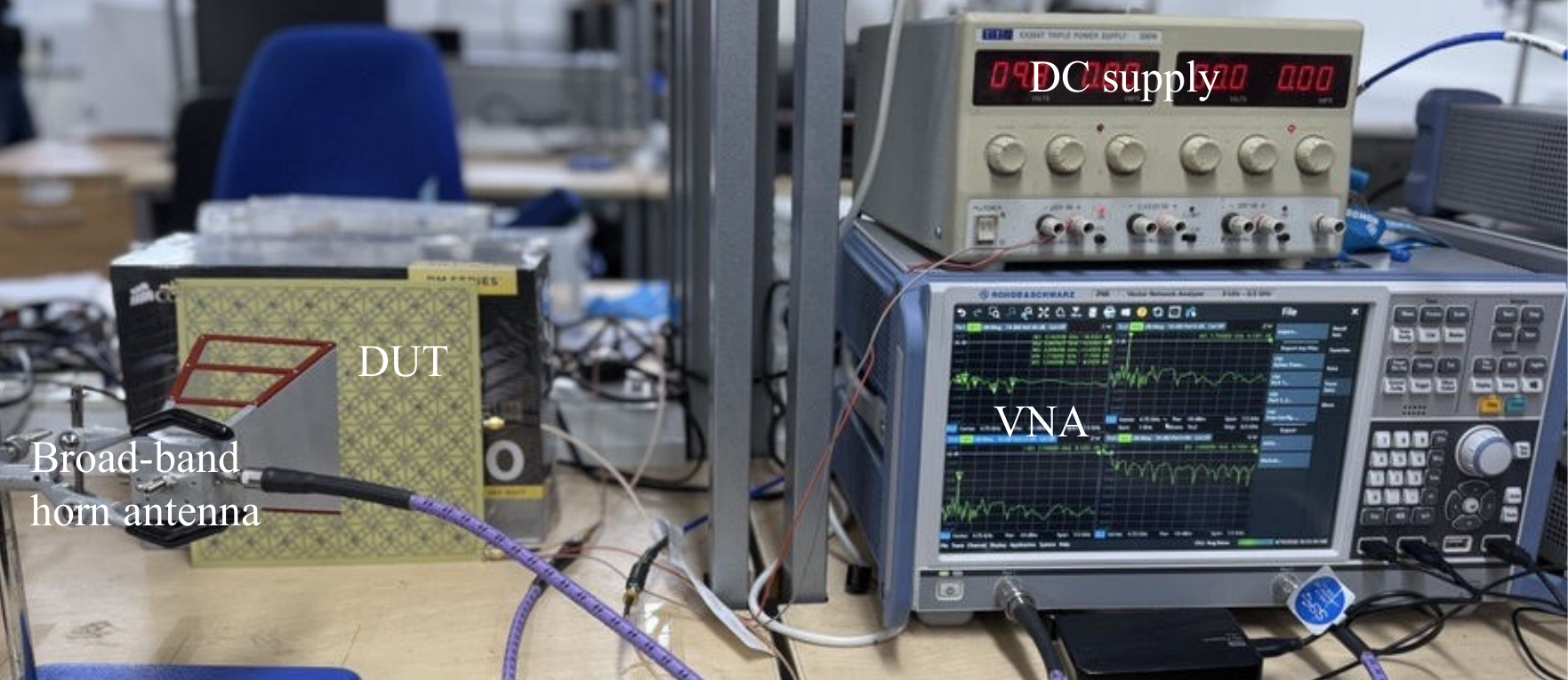}\\[4pt]
	\includegraphics[width=1\columnwidth]{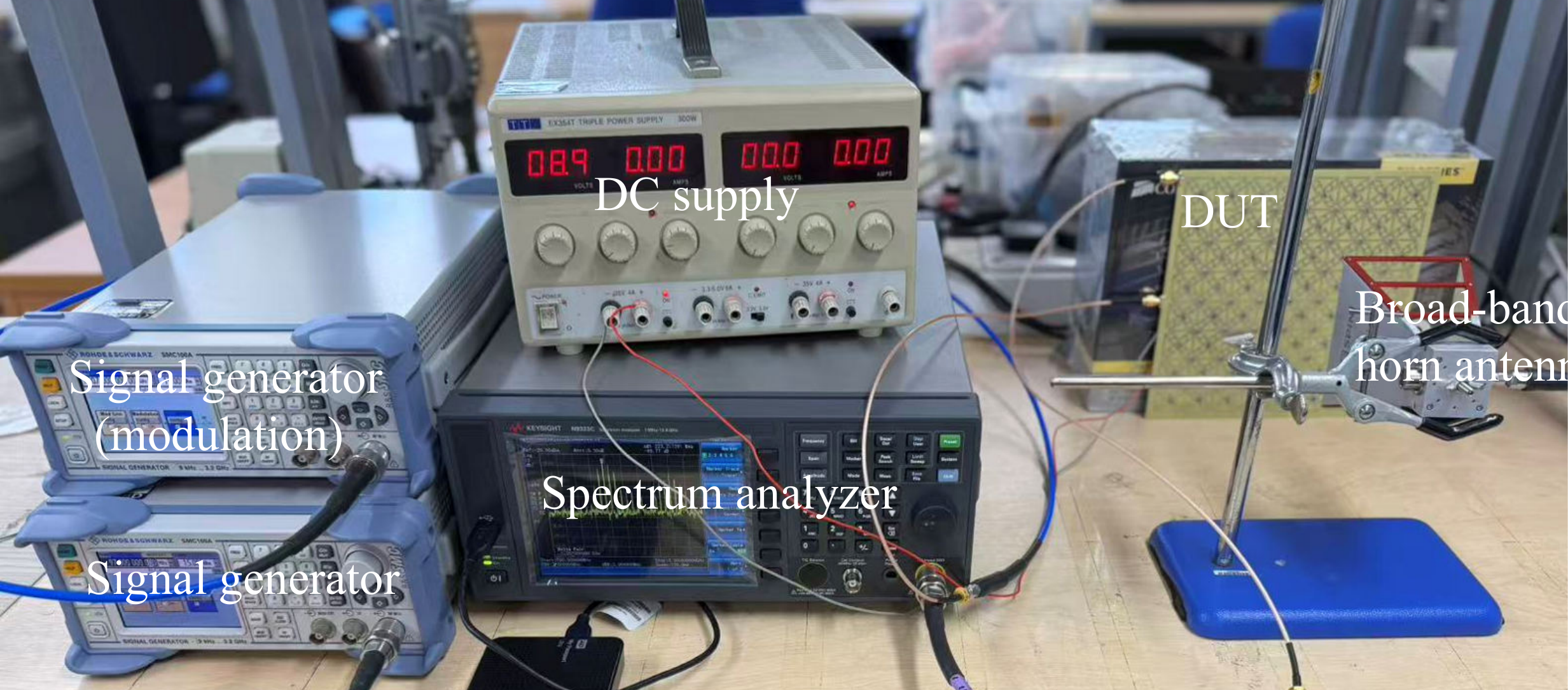}
	\caption{Measurement set-up. (a)~Measurement of the frequency response of the array using a vector network analyser. (b)~Measurement of the frequency generation response of the modulated array using two signal generators and a spectrum analyser.}
	\label{Fig:Resbb}
\end{figure}

Static characterization used a vector network analyzer to record the reflection and transmission coefficients of the unmodulated array (Fig.~\ref{Fig:Resbb}a). Dynamic characterization used two phase-locked signal generators to supply the RF excitation at $f_0$ (TX port, $P_0=-20$~dBm) and the modulation signal at $f_\text{m}$ with controllable relative phase ($\phi_c$ or $\phi_{\rm br}$); the radiated output was captured with a spectrum analyzer and a broadband horn antenna positioned at variable angles $\theta_r$ in an anechoic chamber to map the far-field radiation pattern of each Floquet harmonic (Fig.~\ref{Fig:Resbb}b). For the reception leg, a signal generator and a broadband transmit antenna illuminated the array at $f_1$ with the modulation running, and the power delivered to each port was read directly with a spectrum analyzer.

A standard vector network analyzer computes $S$-parameters as a ratio of reflected/transmitted to incident power under the implicit assumption that the device under test is linear and time-invariant, so that a single stimulus frequency fully determines the response. A time-modulated, polychromatic network such as ours violates this assumption by design: when the probe frequency coincides with the modulation frequency or one of its harmonics, leakage of the modulation tone into the RF path can alias into the same frequency bin as the measured response, and a VNA cannot distinguish this from genuine transmission. We found exactly this behaviour in preliminary $S$-parameter sweeps of the modulated array, including an apparent, frequency-comb-like pattern of both gain and nonreciprocity that we traced to this effect: it reproduced identically on an unrelated, time-modulated test structure with no quadrupole topology, and vanished when the modulation was switched off, confirming it as an artifact of the measurement rather than a property of the device. For this reason, all nonreciprocity and conversion-efficiency results reported here are based on absolute-power measurements referenced to an independent signal generator and spectrum analyzer, rather than on swept $S$-parameter data, for any configuration in which the probe and modulation frequencies could coincide.

The nonreciprocal, frequency-converting transmission was modeled via coherent superposition of the tight-binding propagation paths through the unit cell, \eqref{eq:superposition}. The Floquet spectrum of the driven lattice was checked numerically (Sec.~\ref{sec:floquetcheck}).

A prototype array (Fig.~2 of the main text and Fig.~\ref{Fig:Resba}) implementing the plaquette of \eqref{eq:H0} with integrated varactors and dipoles was fabricated on a microwave substrate, with the two varactors per cell independently biased to set $\phi_c$ and $\phi_{\rm br}$ (Fig.~2 of the main text). A vector network analyzer first characterized the unmodulated array, confirming the static $\pi$-flux quadrupole baseline (Fig.~\ref{Fig:Resbb}a). The array was then driven at $\Omega$ while TX port was excited at $f_0$; the radiated harmonics were captured with a spectrum analyzer and a broadband horn antenna swept in an anechoic chamber (Fig.~\ref{Fig:Resbb}b).

\section{Experimental results: full description}
We first tested the up-conversion mechanism of \eqref{eq:etaup} in the regime $f_\text{m}\gg f_0$, where the modulation frequency far exceeds the injected baseband tone. TX port was excited at $f_0=90$~MHz while the array was driven at $f_\text{m}=2.346$~GHz (Fig.~3 of the main text). The injected tone is spectrally pure at $f_0$ (Fig.~3(a) of the main text), yet the radiated far field is dominated not by $f_0$ but by the up-converted Floquet sideband $f_{+1}=f_0+f_\text{m}=2.436$~GHz at broadside, $\theta_r=0^\circ$ (Fig.~3(b) of the main text), consistent with the resonance-matching argument: $f_0$ sits far below the array's own operating frequency, so it is the $m=+1$ sideband, coinciding with that resonance, that couples efficiently and radiates. The individual Floquet harmonics of \eqref{eq:harmonics} again peak at distinct angles in the normalized radiation pattern (Fig.~3(c) of the main text), confirming that the angular and spectral separation of the corner-state comb persists under this large frequency translation.

We next tested the complementary regime, $f_\text{m}\ll f_0$, exciting TX port directly near the corner mode's own operating frequency, $f_0=2.436$~GHz, while modulating at only $f_\text{m}=90$~MHz (Fig.~4 of the main text). Although the TX excitation is itself spectrally pure (Fig.~4(b) of the main text), the radiated field still resolves a full Floquet comb of sidebands spaced by $f_\text{m}$ (Fig.~4(c) of the main text), with each sideband exhibiting its own distinct E-plane and H-plane radiation pattern (Fig.~4(a) of the main text); the strength of this sideband generation grows with the DC bias applied to the varactors, from 1.5 to 4.1~V (Fig.~4(c) of the main text), confirming that it is a tunable, modulation-driven effect rather than a fixed linear response. Notably, the up-converted radiation in this near-resonance regime and the large-translation up-conversion of Fig.~3(b) of the main text converge on the \emph{same} frequency, $\approx2.436$~GHz, despite being reached via modulation frequencies that differ by more than an order of magnitude ($f_\text{m}=90$~MHz versus $2.346$~GHz); this regime-independent convergence indicates that $\approx2.436$~GHz is a property of a resonance of the structure, rather than an artifact of either particular choice of $f_0$ or $f_\text{m}$.

To test the reception leg independently within this same near-resonance configuration, the array was illuminated from free space at $f_1=2.436$~GHz using a signal generator and a broadband transmit antenna, and the absolute power delivered to TX port and to RX port was measured directly with a spectrum analyzer. The illuminating tone is itself spectrally pure, with no significant frequency generation at the source (Fig.~4(d) of the main text); at RX port this same tone is received strongly and without further conversion (Fig.~4(e) of the main text), while under identical illumination TX port receives a markedly weaker signal (Fig.~4(f) of the main text), with the power delivered to TX port more than $21$~dB weaker than that delivered to RX port. This is consistent with the nonreciprocal reception predicted by \eqref{eq:etaRX} and broken time-reversal symmetry: the incident wave at $f_1$ is preferentially routed to RX port, at the same frequency at which it arrived, while the reverse path to TX port is strongly isolated.
\section{Numerical Floquet check of the corner modes}\label{sec:floquetcheck}
We checked the Floquet spectrum of the drive of \eqref{eq:Hk} with $\gamma\to\gamma(t)$ and $\lambda\to\lambda(t)$ (Eq.~(2) of the main text) on an open lattice of $N\times N$ four-site cells, built from the real-space form of $H(\mathbf{k})$ with the gamma matrices of Ref.~\cite{Benalcazar2017a}. The couplings are $\gamma_0=35$, $\lambda_0=150$, $\delta\gamma=10$ and $\delta\lambda=20$ in units of $2\pi\times$MHz, with $\Omega=2\pi f_\text{m}$. The Floquet operator $U(T)$ was built from 900 to 1500 piecewise-constant steps with exact exponentials, and the quasienergies are $\varepsilon=i\log(\text{eig}\,U)/T$ folded into $(-\Omega/2,\Omega/2]$. The corner weight of a state is its fraction of weight on the four corner cells. This is the idealized lattice model of Eq.~(2), without loss, varactor detail or the full device geometry, and the corner-weight threshold is a choice.

\emph{Zero-quasienergy modes.} For $N=6$, four states at $\varepsilon\approx0$ persist under the drive at every modulation frequency tested (Table~\ref{tab:floquet}).

\emph{No $\pi$-quasienergy corner modes.} At $ f_\text{m}=2346$~MHz and $90$~MHz there are no corner states near $\varepsilon=\pm\Omega/2$. At $ f_\text{m}=90$~MHz the bulk Floquet bands fold onto each other, with 16 bulk states within 10\% of $\varepsilon=0$ and 8 within 10\% of $\pm\Omega/2$ (of 100), so there is no clean gap. Near $ f_\text{m}=300$~MHz eight states with corner weight $0.53$ to $0.59$ appear near $|\varepsilon|\approx145$. They are not $\pi$ modes. The undriven lattice already has eight states at $E=\pm144.75$ and eight at $E=\pm165.43$ with corner weight $0.33$ to $0.39$, and the driven states stay at $|\varepsilon|\approx145$ while $\Omega/2$ is varied. They approach $|\varepsilon|=\Omega/2$ only where $\Omega/2$ crosses the static energies, at $ f_\text{m}\approx289.5$ and $331$~MHz. For $N=5$, a scan over $ f_\text{m}=60$ to $800$~MHz at one, two and three times the modulation depths found no contiguous region with corner states pinned at $\Omega/2$. Two isolated hits (at $60$~MHz with twice the depth and $300$~MHz with three times the depth) were not examined and are consistent with the same crossing effect.

\begin{table}
	\caption{Corner states ($N=6$): number of states with corner weight above $0.5$ at $\varepsilon\approx0$ and near $\pm\Omega/2$.}
	\label{tab:floquet}
	\begin{ruledtabular}
		\begin{tabular}{lccc}
			Case & $\varepsilon\approx0$ & weight & near $\pm\Omega/2$\\
			\hline
			undriven & 4 & 0.89 & 0\\
			$ f_\text{m}=2346$~MHz & 4 & 0.89 & 0\\
			$ f_\text{m}=90$~MHz & 4 & 0.76 & 0\\
			$ f_\text{m}=300$~MHz & 4 & 0.91 & 8 (folded static states)\\
		\end{tabular}
	\end{ruledtabular}
\end{table}

\emph{High-frequency limit.} For four momenta $\mathbf{k}$, the quasienergies of the exact Bloch Hamiltonian $H(\mathbf{k},t)$ differ from the eigenvalues of the time-averaged $H_0(\mathbf{k})$ by at most $0.0094$ at $ f_\text{m}=2346$~MHz and $0.0023$ at $4692$~MHz, in units of $2\pi\times$MHz, against a bandwidth of about 262. The deviation scales as $1/\Omega^2$, so the band structure at $ f_\text{m}=2346$~MHz is that of the static lattice. This check does not cover $ f_\text{m}=90$~MHz, which is comparable to $\lambda_0$.

\bibliography{Taravati_Reference}

\end{document}